\documentclass[a4paper,fleqn,usenatbib]{mnras}

\usepackage{newtxtext,newtxmath}

\usepackage[T1]{fontenc}
\usepackage{ae,aecompl}

\usepackage{graphicx}	
\usepackage{amsmath}	
\usepackage{amssymb}	
\usepackage{bm}

\title[Sink feedback in SPH models of star formation]{Sink particle radiative feedback in smoothed particle hydrodynamics models of star formation}

\author[M. O. Jones et al.]{
Michael O. Jones$^{1}$\thanks{E-mail: mjones@astro.ex.ac.uk (MOJ); mbate@astro.ex.ac.uk (MRB)} and Matthew. R. Bate$^{1}$
\\
$^{1}$School of Physics and Astronomy, University of Exeter, Stocker Road, Exeter EX4 4QL\\
}

\date{Accepted for publication in MNRAS}

\pubyear{2016}

\begin{document}
\label{firstpage}
\pagerange{\pageref{firstpage}--\pageref{lastpage}}
\maketitle

\begin{abstract}
We present a new method for including radiative feedback from sink particles in smoothed particle hydrodynamics simulations of low-mass star formation, and investigate its effects on the formation of small stellar groups. We find that including radiative feedback from sink particles suppresses fragmentation even further than calculations that only include radiative transfer within the gas. This reduces the star-formation rate following the formation of the initial protostars, leading to fewer objects being produced and a lower total stellar mass. The luminosities of sink particles vary due to changes in the accretion rate driven by the dynamics of the cluster gas, leading to different luminosities for protostars of similar mass. Including feedback from sinks also raises the median stellar mass. The median masses of the groups are higher than typically observed values. This may be due to the lack of dynamical interactions and ejections in small groups of protostars compared to those that occur in richer groups. We also find that the temperature distributions in our calculations are in qualitative agreement with recent observations of protostellar heating in Galactic star-forming regions.
\end{abstract}

\begin{keywords}
accretion -- hydrodynamics -- radiative transfer -- stars: formation -- stars: low-mass -- stars: protostars
\end{keywords}



\section{Introduction} \label{intro}
The role of feedback in star formation has received a great deal of attention in recent years. This is largely thanks to the number of studies, both observational and theoretical, showcasing its sometimes dramatic effects. There are many different forms of feedback which apply across a range of different spatial and temporal scales, such as outflows/jets, photoionsation, radiative heating, stellar winds and supernovae. 

Radiative feedback, whilst often ignored on large scales due to the efficient cooling of low density gas, has been shown to be very important in star-formation. Calculations performed by \citet{2007ApJ...656..959K}, \citet{2009MNRAS.392.1363B,2012MNRAS.419.3115B} and \citet{2009ApJ...703..131O} have shown that the additional thermal pressure provided by radiative heating from protostars suppresses fragmentation in systems producing both high- and low-mass stars. This prevents the over-production of brown dwarfs \citep{2009MNRAS.392.1363B,2009ApJ...703..131O}, that was prevalent in earlier calculations that used barotropic equations of state \citep*{2003MNRAS.339..577B,2005MNRAS.356.1201B,2009MNRAS.392..590B}. \citealt{2009MNRAS.392.1363B} and \citealt{2011ApJ...743..110K} have further suggested that radiative feedback may be responsible for the observed invariance of the typical stellar mass in Galactic star formation, by altering the effective thermal Jeans mass of the cloud. 

Most hydrodynamical simulations of star-cluster formation have used sink particles to model protostars \citep{1995MNRAS.277..362B}. In their simplest form, sink particles omit the effects of all physical processes that occur inside the accretion radius, $r_{\text{acc}}$, on the surrounding calculation, barring gravitational forces. However, when mass accretes onto a protostar, its gravitational potential energy is converted into thermal energy, which is then radiated away. As the potential energy released scales inversely with the minimum radius, using sink particles with accretion radii much larger than the typical protostellar radius will significantly underestimate the heating effect of mass accretion. Calculations resolving the formation of stellar cores performed by \citet{2010MNRAS.404L..79B,2011MNRAS.417.2036B} and \citet{2010ApJ...714L..58T} have shown that radiative feedback from the formation of protostars can significantly effect the dynamics of the surrounding envelope, in some cases generating thermally driven outflows.

To include radiative feedback from inside the sink particle, there are two main sources of luminosity which must be accounted for: the intrinsic luminosity of the protostar, and the luminosity generated by the accretion of mass. The intrinsic luminosity of the protostar has been generally ignored in past studies, as for low-mass protostars accreting at high rates, it is expected to be small in comparison with the accretion luminosity (\citealt{2009ApJ...703..131O}; \citealt{2009ApJ...691..823H}). Mass accretion onto a protostar generates luminosity in two ways. The first is compressional heating as the gas collapses or is channelled through an accretion disc. This is radiated by the gas before it reaches the protostar, and accounts for $\approx 50$\% of the total gravitational potential energy released. Most of the remaining gravitational potential energy is released by shock-heating as the accretion flow meets the protostellar surface. Some fraction of the energy is also used in generating jets and outflows, and some will be advected into the protostar to be released on longer timescales.

Several different approaches have been taken in previous studies to model the effects of radiative feedback from accretion. \citet{2007ApJ...656..959K} and \citet{2009ApJ...703..131O} employed sink particles that emitted both the accretion luminosity and that of the central protostar in adaptive mesh refinement calculations. \citet{2010ApJ...710.1343U} approximated the heating effect of protostellar accretion by setting the gas temperature based on the proximity of luminous sources (see \citealt{2009ApJ...698.1341U}). Both of these methods used sink particle accretion radii much larger than the protostellar radius (30 AU for \citealt{2007ApJ...656..959K}, $16-128$~AU for \citealt{2009ApJ...703..131O}; 150 AU for \citealt{2010ApJ...710.1343U}), and consequently they had to make broad assumptions about the dynamics of interior to the sink particles.

\citet{2009MNRAS.392.1363B,2012MNRAS.419.3115B} chose instead to use sink particles with much smaller accretion radii (0.5 AU), and ignore the luminosity generated by material inside the sink. This method has no free parameters other than the accretion radius, and captures much more of the luminosity than is included when using larger ($\sim 100$~AU) sink particles. However, the accretion luminosity is still underestimated by a large factor, $\approx r_{\text{acc}}/R_*$, where $R_*$ is the protostellar radius. \citet{2012MNRAS.419.3115B} argued that the luminosity generated by the resolved gas at small radii was sufficient to suppress most anomalous fragmentation, and that including the remaining luminosity from inside the sink accretion radius would not significantly affect the statistical properties of clusters. Nevertheless, treating the missing luminosity is necessary if we wish to develop more accurate models.

To model the fragmentation of individual discs, \citet*{2011ApJ...730...32S,2012MNRAS.427.1182S} used small sink particles with accretion radii of 1 AU in conjunction with an episodic accretion model, based on the magneto-rotational instability (MRI) driven model of \citet{2010ApJ...713.1134Z}, to provide radiative feedback.  \textbf{Evidence for episodic accretion comes from FU Ori type outbursts from young stars \citet{1996ARA&A..34..207H} and it can be used to help solve the luminosity problem whereby protostars appear to be less luminous than they are expected to be if they accrete at a rate given by their mass over their age \citet{1990AJ.....99..869K}.}  The \citeauthor{2011ApJ...730...32S} model produces short-lived bursts of high accretion rates and large luminosities ($\sim 5 \times 10^{-4}~\text{M}_{\odot}~\text{yr}^{-1}$; $\sim 10^3~\text{L}_{\odot}$) lasting a few hundred years, followed by long quiescent periods of low accretion rates and almost negligible luminosities ($\sim 10^{-7}~\text{M}_{\odot}~\text{yr}^{-1}$; $\sim 0.1~\text{L}_{\odot}$) lasting a few thousand years. Stamatellos et al. argued that these quiescent periods are essential for correctly modelling the production of brown dwarfs, by allowing circumstellar discs to cool and fragment into low-mass objects. This model was later used by \citet{2014MNRAS.439.3039L,2015MNRAS.447.1550L} to study star formation in small stellar groups.  Although the model only has a few free parameters, it relies on the specific MRI-driven episodic accretion model and, observationally, accretion burst luminosities, durations, and repeat times are highly uncertain.

In this paper we introduce a method, based on continuous disc accretion onto a central protostar, for including the accretion luminosity generated inside sink particles in SPH calculations of star-formation. We combine our method with sink particles of the same size as previously used by \citet{2009MNRAS.392.1363B}. This enables us to more accurately include the majority of the radiative feedback produced by protostars whilst making relatively few assumptions.

Section \ref{method} describes the numerical method, including the model used to calculate the radiative heating from sink particles, and the initial conditions used for each calculation. Section \ref{resultsBE} presents results from radiative hydrodynamics simulations of the formation of single protostellar systems using the new model to calculate accretion luminosities. In Section \ref{resultsclus}, we analyse the effects of the extra feedback on simulations of the formation of small, ($50~\text{M}_{\odot}$) stellar groups, using similar initial conditions to those used by \citet{2009MNRAS.392.1363B}. Section \ref{disc} compares our results to those obtained from previous attempts to include radiative feedback from sink particles, and also to observations of radiative feedback. Our conclusions are given in Section \ref{conc}.

\section{Computational Method} \label{method}
The calculations presented in this paper were completed using \texttt{sphNG}, a modified version (\citealt{1995MNRAS.277..362B}, \citealt{2005MNRAS.364.1367W}, \citealt{2006MNRAS.367...32W}) of a three-dimensional smoothed particle hydrodynamics (SPH) code originally developed by \citet{1990ApJ...348..647B,1990nmns.work..269B}, parallelised using both \texttt{OpenMP} and \texttt{MPI}. The code uses a binary tree to calculate the gravitational forces between particles and their nearest neighbours. The smoothing lengths of particles are allowed to vary in time and space, and are set iteratively such that the smoothing length of each particle $h = 1.2(m/\rho)^{1/3}$ where $m$ is the particle mass, and $\rho$ is the particle density (see \citealt{2007MNRAS.374.1347P}). A second-order Runge-Kutta-Fehlberg method \citep{1969NASA...R315} is used to integrate the SPH equations, with individual time-steps for each particle \citep{1995MNRAS.277..362B}. The artificial viscosity prescription given by \citet{1997JCoPh.136...41M} is used, with $\alpha_v$ varying between 0.1 and 1 and $\beta_v = 2\alpha_v$.

\subsection{Radiative transfer \& equation of state} \label{method_RHD}
In the co-moving frame, assuming local thermal equilibrium (LTE), the frequency-integrated equations describing the time-evolution of radiation hydrodynamics (RHD) are
\begin{equation}
	\frac{\text{D}\rho}{\text{D}t}  = \rho \nabla \cdot \bm{v} = 0,
\end{equation}
\begin{equation}
	\rho \frac{\text{D}\mathbf{v}}{\text{D}t} = -\nabla p + \frac{\kappa \rho}{c} \bm{F},
\end{equation}
\begin{equation} \label{eq:radiationenergy}
	\rho \frac{\text{D}}{\text{D}t} \left(\frac{E}{\rho}\right) = -\nabla \cdot \bm{F} - \nabla \bm{v} : \mathbf{P} + 4\pi \kappa \rho B - c \kappa \rho E + \rho \Gamma_*, 
\end{equation}
\begin{equation} \label{eq:matterenergy}
	\rho \frac{\text{D}}{\text{D}t} \left(\frac{e}{\rho}\right) = -p \nabla \cdot \bm{v} - 4\pi \kappa \rho B + c \kappa \rho E, 
\end{equation}
\begin{equation}
	\frac{\rho}{c^2} \frac{\text{D}}{\text{D}t} \frac{\bm{F}}{\rho} = -\nabla \cdot \mathbf{P} - \frac{\kappa \rho}{c} \bm{F}, 
\end{equation}
\citep{1984oup..book.....M}, where $\text{D}/\text{D}t = \partial / \partial t + \bm{v} \cdot \nabla$ is the convective derivative. $\rho, e, \bm{v}$ and $p$ are the mass density, energy density, velocity and scalar isotropic pressure, and $E$, $\bm{F}$ and $\mathbf{P}$ represent the frequency-integrated radiation energy density, flux and pressure tensor. $B$ is the frequency-integrated Planck function, and $\kappa$ is the opacity.  The last term in equation \ref{eq:radiationenergy}, $\rho \Gamma_*$, is a source term to input radiative feedback from sink particles and is discussed in detail below.

To close this system of equations, we choose an ideal gas equation of state  $p = \rho T_g \mathcal{R} / \mu$, where $T_g$ is the gas temperature, $\mathcal{R}$ is the gas constant and $\mu$ is the mean molecular weight of the gas, set to $\mu = 2.38$. Translational, rotational and vibrational degrees of freedom of molecular hydrogen are accounted for in the thermal evolution of the gas, as well as molecular dissociation of hydrogen and ionisation of both hydrogen and helium, with hydrogen and helium mass fractions of $X = 0.70$ and $Y = 0.28$. We ignore the contribution of metals to the equation of state. 

We implement two-temperature (gas and radiation) radiative transfer using the flux-limited diffusion approximation, as described by \citet{2005MNRAS.364.1367W} and \citet{2006MNRAS.367...32W}, to model the transport of radiation. Energy is generated by doing work on the gas or radiation field, and transferred between the two according to their relative temperatures as well as the gas density and opacity. We assume that the gas and dust are well-coupled, and that their respective temperatures are the same. A grey opacity is used, set to maximum value of the interstellar grain opacity according to the tables of \citet{1985Icar...64..471P} for low temperatures, and the gas opacity for solar metallicity gas according to the tables of \citet{2005ApJ...623..585F} at high temperatures. 

The molecular clouds have free boundaries. However, to provide a boundary for the radiation field, all SPH particles with a density less than $10^{-21}\text{~g~cm}^{-3}$ have both their gas and radiation temperatures set to 10 K.

\subsection{Sink particles} \label{method_sink}
Using radiation hydrodynamics with a realistic equation of state allows us to capture each phase of protostar formation \citep{1969MNRAS.145..271L}. However, as the first hydrostatic core collapses following the dissociation of molecular hydrogen, the time-step as defined by the Courant-Friedrichs-Levy condition becomes very small, making the required calculation time unfeasibly long, particularly for larger calculations involving multiple protostars.

To alleviate this issue, we use sink particles as introduced by \citet{1995MNRAS.277..362B}. Once a particle reaches a density of $\rho_{\text{crit}} = 10^{-5}\text{~g~cm}^{-3}$, it is replaced with a sink particle. The sink particles accrete SPH particles that pass within the accretion radius and the mass of any accreted particles is subsequently added to the sink mass. For the Bonnor-Ebert sphere calculations, we use sink accretion radii of 0.05 AU, 0.5 AU, 5 AU and 50 AU, and for the cluster calculations we use accretion radii of 0.5 AU and 5 AU. Sink particles are permitted to merge if they pass within 0.015 AU of each other, as young protostars are expected to be larger than the Sun (e.g. \citealt{2009ApJ...691..823H}). The sink particles we use do not attempt to correct for the discontinuous pressure and viscous forces at the accretion radius, although several methods have been proposed for doing so \citep[e.g.][]{1995MNRAS.277..362B,2013MNRAS.430.3261H}.

\subsubsection{Sink particle luminosities} \label{method_luminosity}
To model the accretion luminosity generated inside in each sink particle, we use a `sub-grid' model based on the accretion of mass through a circumstellar accretion disc onto a central protostar. Due to conservation of angular momentum, infalling gas will generally form an accretion disc inside the sink particle accretion radius. The protostar itself accretes from this disc at a rate that, in general, will differ from the rate at which the sink particle accretes gas through its accretion radius. 

To describe this, we decompose the total sink mass into a `protostellar mass' and a `disc mass', in a similar manner to \citet{2011ApJ...730...32S,2012MNRAS.427.1182S}. Mass from gas particles accreted by the sink particle is initially stored in the disc. It is then allowed to accrete onto the central protostar from the disc at a rate given by
\begin{equation} \label{discacc}
	\dot{M}_{*} = \dot{M}_0 \left( \frac{M_{\text{D}}}{M_{\text{D},0}} \right) ~\text{M}_{\odot}~\text{yr}^{-1},
\end{equation}
where $M_{\text{D}}$ is the current mass of the disc, and the initial mass of the disc, $M_{\text{D},0}$, is the difference between the initial mass of the sink particle and the initial mass of the protostar, i.e. $M_{\text{D},0} \equiv M_{\text{sink},0} - M_{*,0}$. In the absence of accretion by the sink particle, equation \ref{discacc} results in the protostar accreting mass from the disc at an exponentially decreasing rate. This inherently produces a smoothly changing accretion rate, rather than producing bursts. There are two free parameters: the initial protostellar mass, $M_{*,0}$, and the initial accretion rate, $\dot{M}_0$. The initial protostar is assumed to be an object with a mass of $1\text{M}_{\text{J}}$, and a radius of $2\text{R}_{\odot}$ \citep{1969MNRAS.145..271L}. We set the initial accretion rate to $\dot{M}_0 = 10^{-5}~\text{M}_{\odot}~\text{yr}^{-1}$, since previous calculations have shown that this is a typical value for newly formed protostars \citep{2010MNRAS.404L..79B,2011MNRAS.417.2036B} and for young protostars in clusters \citep{2005MNRAS.356.1201B,2009MNRAS.392..590B,2009MNRAS.392.1363B}. 

The accretion luminosity is calculated from the mass accretion rate as
\begin{equation}
	L_{\rm acc} = \epsilon \frac{G M_{*} \dot{M}_{*}}{R_{*}},
\end{equation}
where $M_*$ is the protostellar mass, and $G$ is the gravitational constant.  For simplicity, we set the efficiency factor for the conversion of potential energy to luminosity, $\epsilon = 1$.  Note that this provides the maximum possible feedback; some potential energy maybe converted to kinetic energy (e.g. jets), or thermal energy that is advected into the protostar to be released on longer timescales. For example, \cite{2009ApJ...703..131O} use $\epsilon = 0.75$.

Previous calculations by \citet{2010MNRAS.404L..79B} that resolved protostellar core formation showed that newly formed protostars undergo an initial phase of very rapid accretion ($\geq 10^{-3}~\text{M}_{\odot}~\text{yr}^{-1}$), lasting approximately 10 years, as the remnants of the first hydrostatic core are accreted. This is also true for calculations using sink particles to model protostars. During this time, the properties of the protostar and its surrounding material undergo rapid changes and a star with disc is a poor representation of the system. We therefore delay initialisation of the model and the emitted luminosity until 10 years after the sink particle forms.

The mass of the protostar is allowed to evolve according to the calculated accretion rate.  We keep the protostellar radius constant throughout the calculations.  This is a reasonable approximation for low-mass protostars ($M_* \lesssim 1~{\rm M}_\odot$) accreting at rates of $\dot{M}_* \approx 10^{-6}-10^{-5}$~M$_\odot$~yr$^{-1}$, as shown by the accreting stellar models of \citet{2009ApJ...691..823H}.  We choose to ignore the protostar's intrinsic luminosity, as the accretion luminosity dominates over the intrinsic luminosity when the protostars are undergoing significant accretion (\citealt{2009ApJ...703..131O}; \citealt{2009ApJ...691..823H}).  It would be straightforward to incorporate more complicated protostellar evolutionary models to compute the combined accretion and intrinsic luminosity, but this would introduce more assumptions and parameters and this additional complexity is beyond the scope of this paper.  Our aim here is to determine the predominant effects of including the luminosity that was missing in the past radiation hydrodynamical SPH models of low-mass star formation.

\subsubsection{Adding the radiative feedback into the simulation} \label{method_feedback}
The protostellar luminosity determined from the above model must be radiated from the sink particle into the SPH simulation. This is done by including the source term $\rho \Gamma_*$ in the righthand side of equation \ref{eq:radiationenergy}, where $\Gamma_*$ has units of $\text{erg}~\text{s}^{-1}~\text{g}^{-1}$.

The source term for a particular SPH gas particle due to the radiative feedback from an individual sink particle is determined as follows:

First, the gas particles surrounding the sink particle are examined to determine which are directly exposed to the radiation emitted by the sink particle.  This is implemented by walking the tree structure that is also used to compute neighbouring particles and gravitational forces. A particle is determining to be exposed if no other gas particle `blocks' the radiation from the sink particle.  If there is another gas particle that is closer to the sink particle than the gas particle that is being examined, and the line segment from the sink particle's position to the position of the gas particle that is being examined intersects the smoothing sphere of the closer gas particle, then the gas particle is considered to be `blocked'.  The smoothing sphere of an SPH gas particle has a radius $r_{\rm W}h$, where $h$ is the SPH smoothing length and $r_{\rm W}$ is the number of smoothing lengths required for the SPH kernel to drop to zero (i.e. for the standard cubic spline kernel $r_{\rm W}=2$). An additional requirement for the particle to be blocked is that the distance between the two gas particles must be greater than the smoothing radius ($r_{\rm W}h$) of the intervening gas particle, otherwise both are considered to be potentially exposed.  

Once a list of exposed particles has been constructed, the fraction of the sink particle's total luminosity that is distributed to each particle must be determined.  This is done by calculating the solid angle subtended by the gas particle with smoothing length $h_i$ with respect to the sink particle which we compute as
\begin{equation}
	\Omega_i =  2 \pi \left(1-\cos(\sin^{-1}(\min(r_{\rm W}h_i,r)/r))\right),
\end{equation}
where $r$ is the distance between the sink particle and the gas particle.  These values are then normalised such that the sum of the solid angles for all exposed particles is unity (i.e. all of the sink particle luminosity is distributed to the SPH particles).  The source term for gas particle, $i$, due to the feedback from a particular sink particle is given by
\begin{equation}
	 \Gamma_{*i}  = \frac{L_* \Omega_i}{m_i \sum_j \Omega_j},
\end{equation}
where $m_i$ is the mass of the gas particle, and the sum over $j$ is done over all of the exposed particles for that sink particle.

The above algorithm is sufficient for a calculation that uses a global time steps for all particles.  However, \texttt{sphNG} uses individual time steps such that particles have time-steps that may differ by powers of two \citep{1995MNRAS.277..362B}.  In this case, only those particles that are currently being evolved have their solid angles re-computed, and the sum of the solid angles over all exposed particles will not always equal unity.  Furthermore, some particles with short time steps may be accreted between the time steps of exposed particles with longer time steps.  To adjust for this, the amount of energy actually being added into the simulation over the time step of the exposed particle with the longest time step is calculated and if this energy is slightly too large or small this deficit or excess is corrected over the subsequent period of time equal to the length of the time step of the exposed particle with the longest time step.

The above method of distributing the radiant energy to surrounding SPH gas particles essentially assumes that the surrounding particles are optically thick and the radiation is absorbed close to the protostar.  This will be the case for rapidly accreting, young protostars.  If a sink particle is not surrounded by optically thick gas, the radiant energy emitted by the sink particle is still added into the radiation hydrodynamical simulation, but it will then propagate to larger distances as prescribed by flux-limited diffusion with all the usual limitations of that approximation (in particular, it will propagate diffusively rather than along rays).

Equations \ref{eq:radiationenergy} and \ref{eq:matterenergy} are solved using the same iterative, implicit method as that described by \citet{2005MNRAS.364.1367W} and \citet{2006MNRAS.367...32W}.  The only differences are the presence of the extra source term, $\rho \Gamma_*$, and the fact that rather than use an implicit expression for the $p\nabla\cdot \mbox{\bm{$v$}}$ term in equation \ref{eq:matterenergy}, since \citet{2010MNRAS.404L..79B} an explicit expression has been used for better energy conservation \citep[see Section 3.1 of ][]{2015MNRAS.449.2643B}.

\begin{figure*}
	\includegraphics{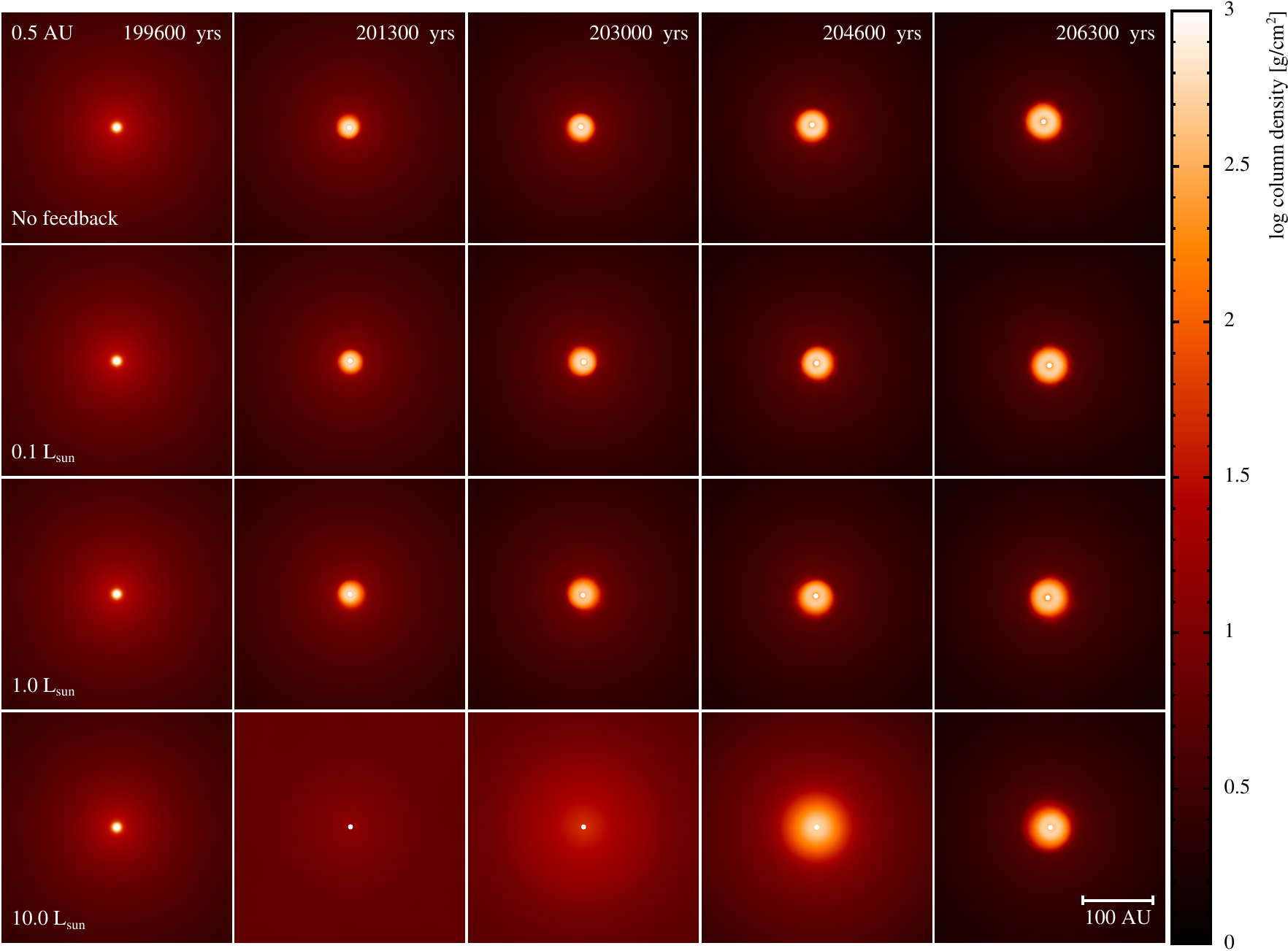}
	\caption{The time evolution of the gas column density in each of the Bonnor-Ebert sphere calculations, using sink particles with accretion radii of 0.5 AU and constant luminosities. The luminosities used in the calculations are given in the first panel of each row. We also plot results from a calculation using a sink particle with no luminosity for comparison (top row). The panels display the calculations at times separated by intervals of 1,657 yr, which corresponds to 1/200th of the initial cloud free-fall time. The colour scale shows column densities on a logarithmic scale between $1~\text{g~cm}^{-2}$ and $1000~\text{g~cm}^{-2}$, viewed parallel to the cloud's rotation axis. The length scale is shown in the bottom-right of the plot. The constant luminosity of the sink particle has little effect on the system for luminosities of $0.1 \text{L}_{\odot}$ and $1 \text{L}_{\odot}$. However, using a luminosity of $10 \text{L}_{\odot}$ rapidly heats the inner disc, driving an expansion that temporarily destroys the disc.}
	\label{dens_co_0001_05AU}
\end{figure*}

\begin{figure*}
	\includegraphics{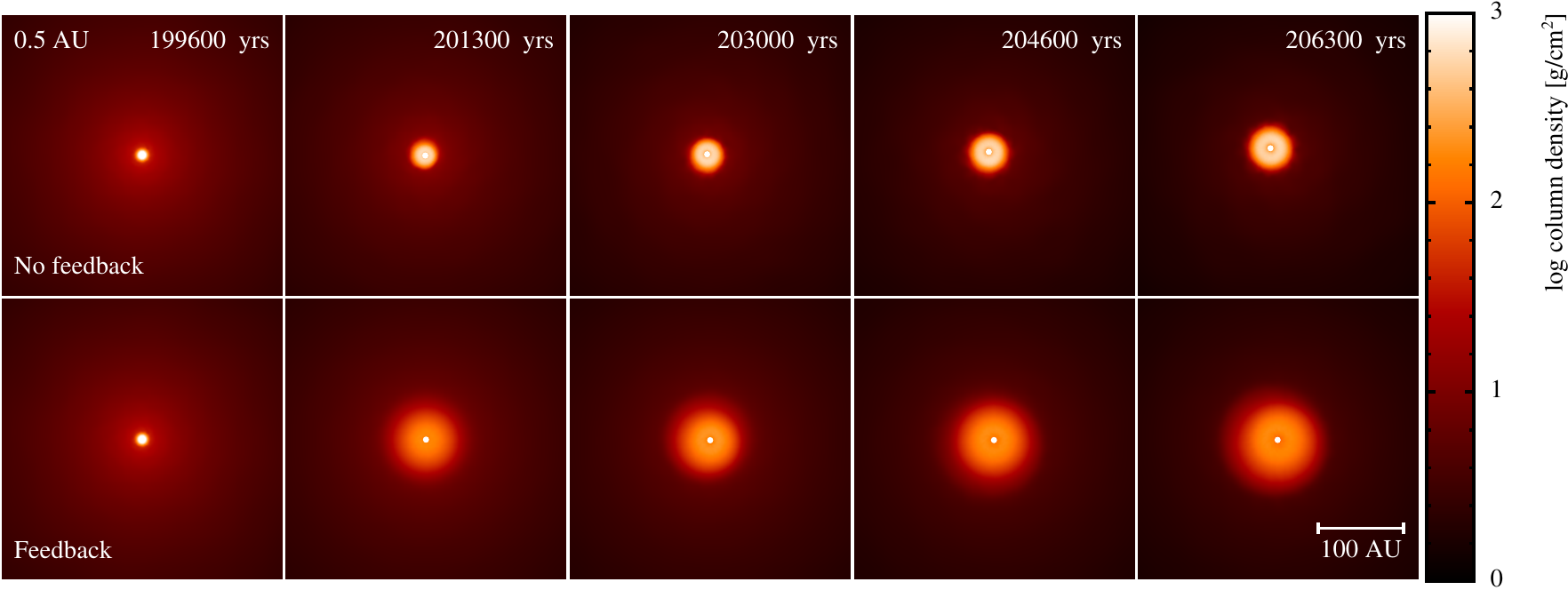}
	\includegraphics{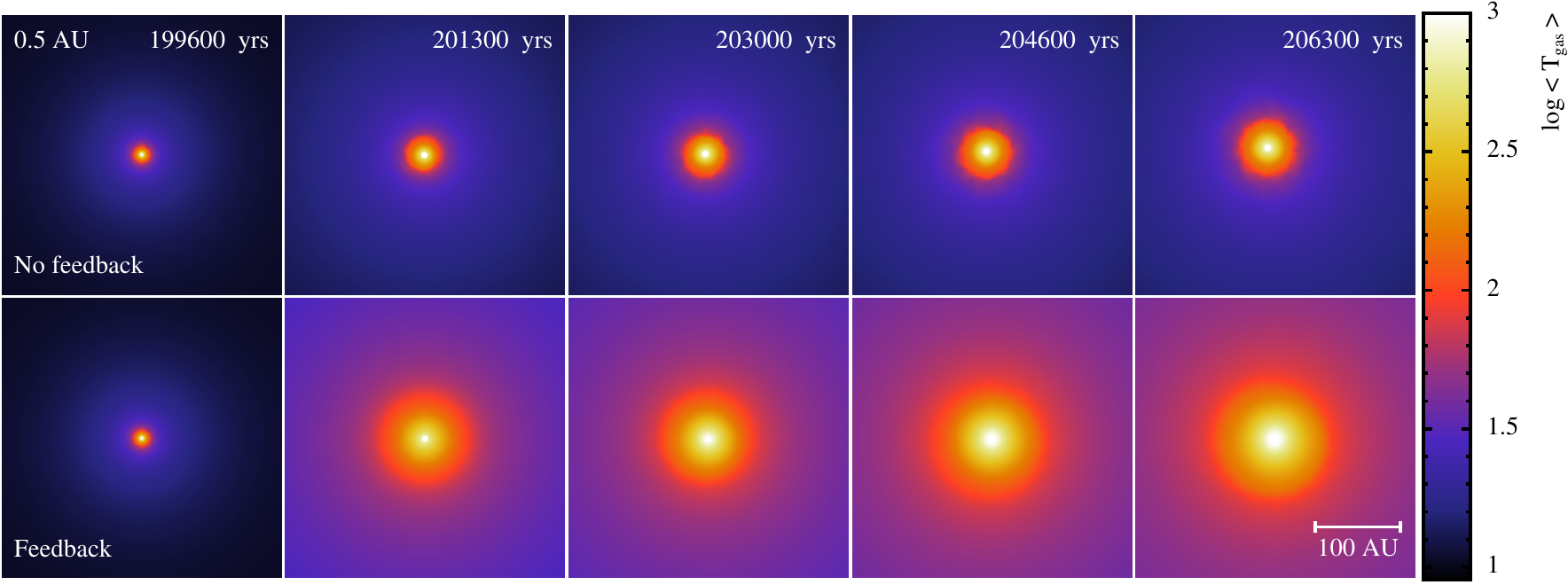}
	\caption{
	The time evolution of the gas column density and mass-weighted temperature in the Bonnor-Ebert sphere calculations, using sink particles with accretion radii of 0.5 AU without feedback (top row of each sub-panel) and with luminosities set by the disc-accretion algorithm (bottom row of each sub-panel). The panels display the calculations at times separated by intervals of 1,657 yr, which corresponds to 1/200th of the initial cloud free-fall time. The colour scales show column densities on a logarithmic scale between $1~\text{g~cm}^{-2}$ and $1000~\text{g~cm}^{-2}$, viewed parallel to the cloud's rotation axis, and temperatures on a logarithmic scale between 9 K and 1000 K. The length scale is shown in the bottom-right of each plot. The increasing luminosity of the sink particle heats the disc, causing it to expand steadily throughout the calculation. The result is a much larger and hotter disc than in the calculation that does not include sink particle feedback.}
	\label{denstemp_fb_0001_05AU}
\end{figure*}

\begin{figure*}
	\includegraphics{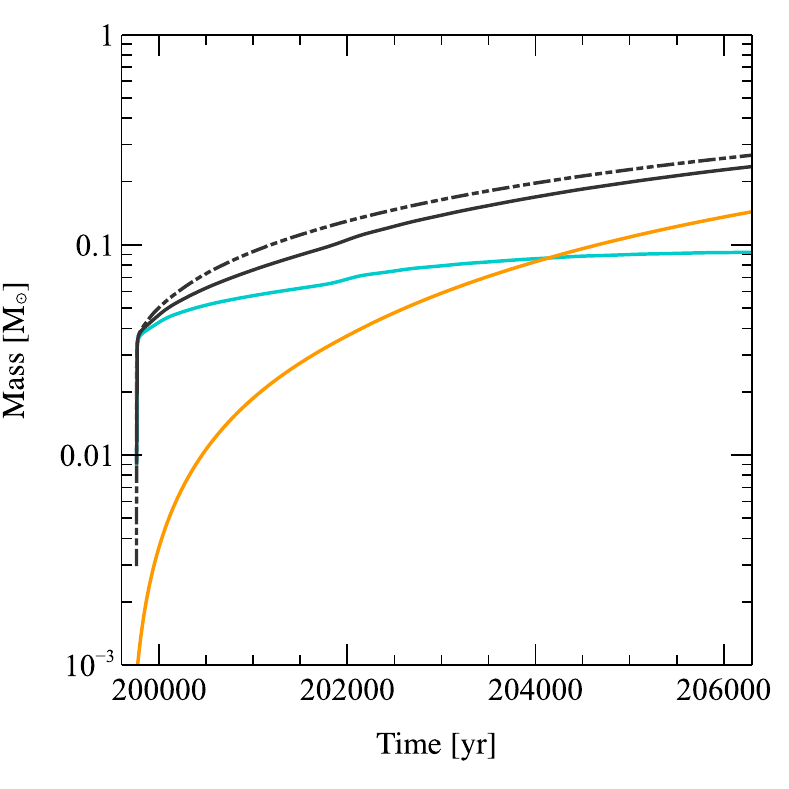}
	\includegraphics{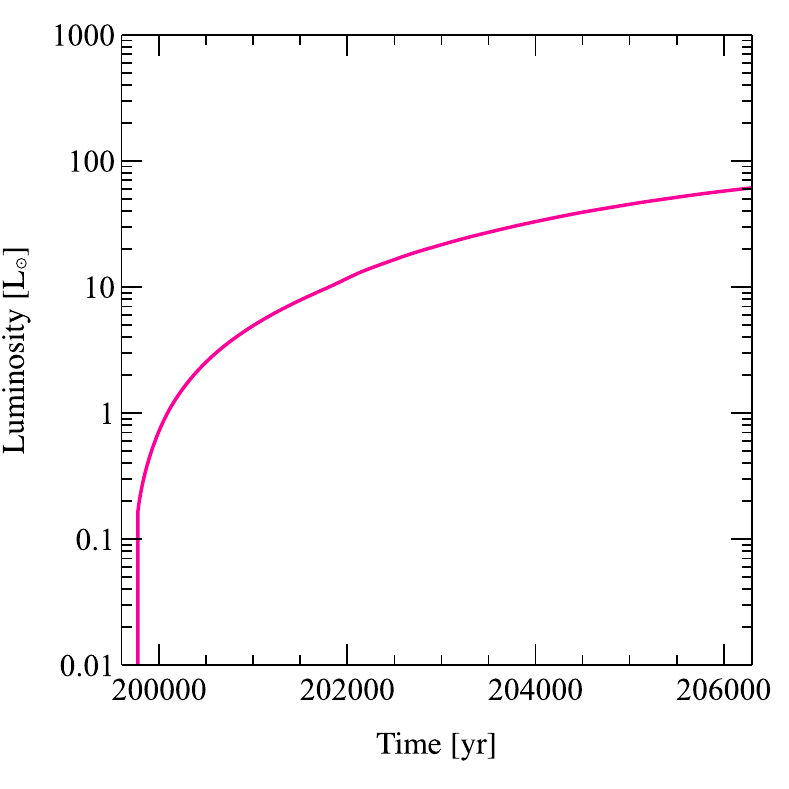}
	\caption{
	The evolution of the sink particle, protostellar, and disc masses and sink luminosities in the Bonnor-Ebert sphere calculations, using sink particles with accretion radii of 0.5 AU without feedback and with luminosities set by the disc-accretion algorithm. In the left plot, the total sink particle masses are shown in black, and the disc and protostellar masses are shown in blue and orange respectively. In the right plot, the sink luminosity is shown in magenta. We also plot the results of a calculation not including radiative sink feedback for comparison, which are shown by the dot-dashed line. Including feedback reduces the mass accretion rate of the sink particle, resulting in a slightly lower sink mass compared to the no-feedback calculation at the same time. The luminosity continues to increase for the duration of the calculation, due to the growth of the stellar mass via accretion.}
	\label{masslumi_fb_0001_05AU}
\end{figure*}

\subsection{Initial conditions} \label{method_initial}
In the calculations presented in Section \ref{resultsBE}, we simulate Bonnor-Ebert spheres containing $5~\text{M}_{\odot}$ of gas, with a uniform initial temperature of 10 K. The spheres have a radius of 0.126 pc, and a centre-to-edge density ratio of 14:1, such that the ratio of thermal energy to gravitational potential energy is $\alpha = 0.4$. The spheres are set in solid body rotation with an angular velocity of $8.02 \times 10^{-14}~\text{rad~s}^{-1}$, such that the ratio of rotational kinetic energy to the magnitude of the gravitational potential energy is $\beta = 0.001$.  Some calculations were also performed with higher rotation rates ($\beta = 0.005$). The initial free-fall time of the spheres, based on their mean density, is $1.05 \times 10^{13}~\text{s}$ ($331,358~\text{yr}$).

For the calculations presented in Section \ref{resultsclus}, we simulate turbulent, uniform density gas clouds, using the same initial conditions as \citet{2009MNRAS.392.1363B}. We construct uniform spheres containing $50~\text{M}_{\odot}$ with uniform initial temperatures of 10.3 K. The spheres have a radius of 0.188 pc and an initial mean density of $1.2 \times 10^{-19}~\text{g~cm}^{-3}$. An initial `supersonic', turbulent velocity field is applied, in the manner described by \citet{2001ApJ...546..980O} and \citet{2003MNRAS.339..577B}. To do this, a divergence-free, random Gaussian field is generated with a power spectrum of $P \propto k^{-4}$, where $k$ is the wavenumber, giving a velocity dispersion which varies with distance $\lambda$ as $\sigma(\lambda) \propto \lambda^{-1/2}$ in three dimensions, in agreement with the Larson scaling relations for molecular clouds \citep{1981MNRAS.194..809L}. The field is generated on a $64^3$ grid and the velocities interpolated from the grid. The field is normalised such that the total kinetic energy is equal to the magnitude of the gravitational potential energy of the sphere. The initial free-fall time of the spheres is  $6.02 \times 10^{12}~\text{s}$ ($190,976 ~\text{yr}$).

\subsection{Resolution} \label{method_res}
To correctly model fragmentation down to the opacity limit, the local Jeans mass must be resolved (\citealt{1997MNRAS.288.1060B}; \citealt{1997ApJ...489L.179T}; \citealt{1998MNRAS.296..442W}; \citealt{2000ApJ...528..325B}; \citealt{2006A&amp;A...450..881H}).  We use $1 \times 10^{6}$ particles to model the $5~\text{M}_{\odot}$ Bonnor-Ebert spheres, and $3.5 \times 10^{6}$ particles to model the $50~\text{M}_{\odot}$ clouds, giving mass resolutions of $1 \times 10^{-6}~\text{M}_{\odot}$ per particle ($2 \times 10^{5}$ particles per $\text{M}_{\odot}$) for the Bonnor-Ebert spheres and $1.4 \times 10^{-5}~\text{M}_{\odot}$ per particle ($7 \times 10^{4}$ particles per $\text{M}_{\odot}$) for the turbulent clouds.

\section{Bonnor-Ebert spheres} \label{resultsBE}
In this section, we present results from radiation hydrodynamical calculations of star formation from the collapse of a single, rotating, Bonnort-Ebert sphere. We examine the effects of including sink particles with constant luminosities of $\text{0.1~L}_{\odot}$, $\text{1.0~L}_{\odot}$ and $\text{10.0~L}_{\odot}$ (Section \ref{resultsBE_cons}), as well as those with luminosities calculated by the disc-accretion algorithm (Section \ref{resultsBE_disc}) using sink particles with accretion radii of $r_{\rm acc}=0.5$~AU. We also compare the results of disc-accretion calculations using accretion radii of 0.05 AU, 0.5 AU, 5 AU and 50 AU (Section \ref{resultsBE_rad}).

The initial phase of cloud collapse is approximately isothermal, due the efficiency of radiative cooling at low densities. This initial phase lasts until the maximum density in the simulation reaches $\sim 10^{-13}~\text{g~cm}^{-3}$, at which point the collapse becomes approximately adiabatic, forming a hydrostatic `first core' (see \citealt{1969MNRAS.145..271L}). Once the temperature reaches $\approx 2000 \text{K}$, molecular hydrogen begins to dissociate, absorbing energy and lowering the opacity, causing the first core to collapse to form a protostar. Once the maximum density of the core reaches $10^{-5}~\text{g}~\text{cm}^3$ (which occurs at a time of 140,343 yr in all of the calculations), a sink particle is inserted.

\subsection{Constant luminosities} \label{resultsBE_cons}
Fig. \ref{dens_co_0001_05AU} shows the evolution of the gas column density in Bonnor-Ebert sphere calculations with an initial rotational energy to gravitational energy ratio of $\beta = 0.001$ using sink particles that have constant luminosities after they are inserted. 

The impact of including the sink luminosity varies between the calculations. In each of the calculations, once the sink is inserted, it begins to heat the surrounding disc material.  If the heating is significant, the resultant increase in pressure may cause the disc to expand. In the $\text{0.1~L}_{\odot}$ and $\text{1.0~L}_{\odot}$ calculations, the heating effect is small, enabling the disc to radiate the additional energy away with little change in its radius. As such, the results are highly similar to the no-feedback case. The heating effect of the $\text{10.0~L}_{\odot}$ sink, however, is strong enough to overcome the weak binding energy of the disc around the newly-formed protostar. This causes the surrounding material to expand dramatically, temporarily destroying the disc and driving a large outflow. The expanding gas eventually cools and re-collapses, forming a disc of similar size and mass to the disc formed in the no-feedback calculation. In calculations with higher rotation rates (e.g. $\beta=0.005$), this process can be unstable to fragmentation if the sink particle becomes displaced from the outflow's centre of mass and has a low mass relative to the outflow mass.  In this case, the dynamics of the collapse are dominated by the gas mass, forming a new object at the outflow's centre of mass. 

\begin{figure*}
	\includegraphics{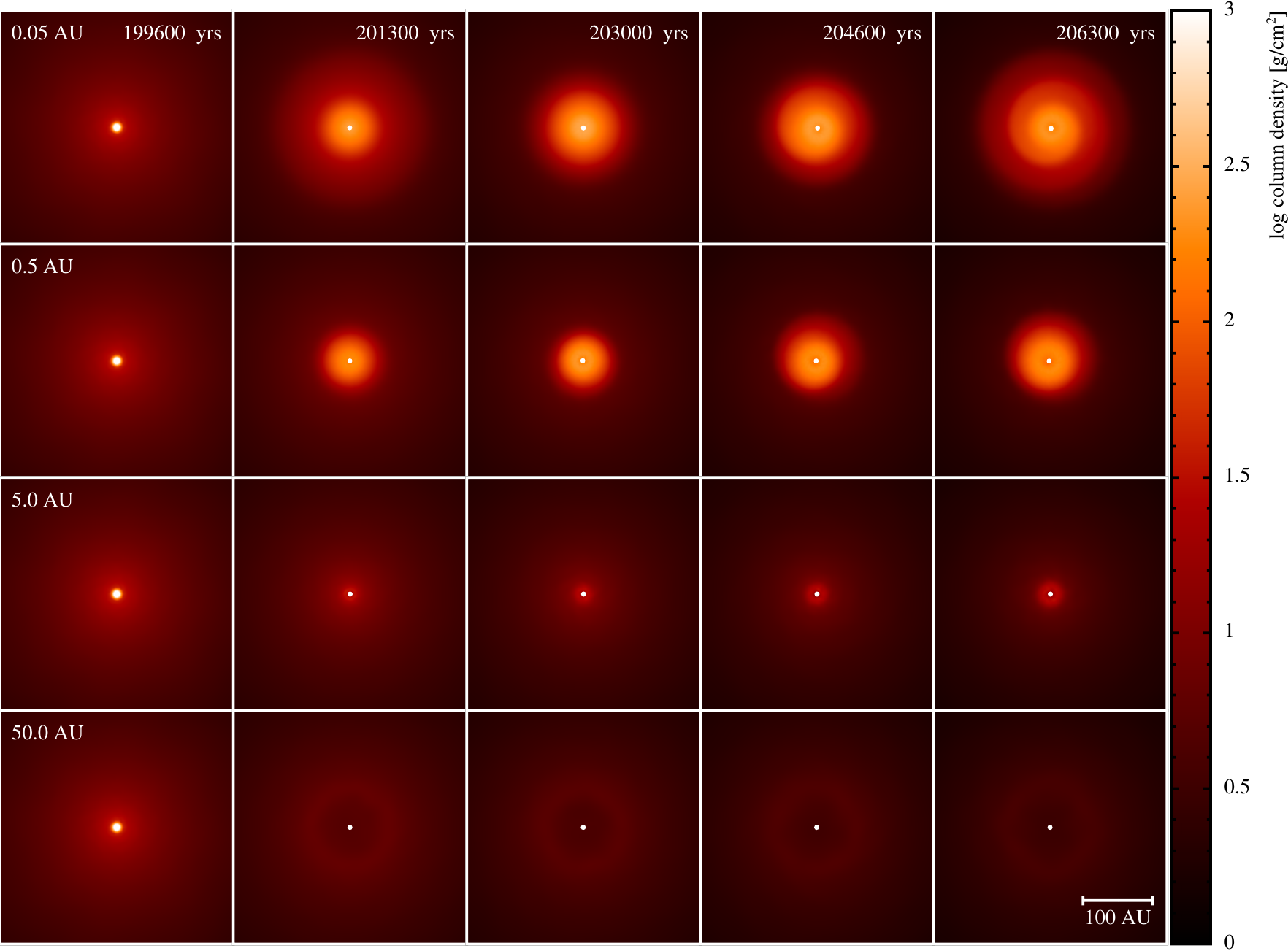}
	\caption{
	The time evolution of the gas column density in the Bonnor-Ebert sphere calculations, using sink particles with luminosities set by the disc-accretion algorithm and a range of accretion radii. The accretion radii used in the calculations are given in the first panel of each row. The panels display the calculations at times separated by intervals of 1,657 yr, which corresponds to 1/200th of the initial cloud free-fall time. The colour scale shows column densities on a logarithmic scale between $1~\text{g~cm}^{-2}$ and $1000~\text{g~cm}^{-2}$, viewed parallel to the cloud's rotation axis. The length scale is shown in the bottom-right of the plot. Using a larger accretion radius binds more material inside the sink particle, preventing it from escaping. In the case of the 0.05-AU sink, the gas expands as an outflow, increasing the disc radius, whereas the 5-AU and 50-AU sinks encompass the majority of the disc material, such that little or no disc is formed.}
	\label{dens_fb_0001_comp}
\end{figure*}

\begin{figure*}
	\includegraphics{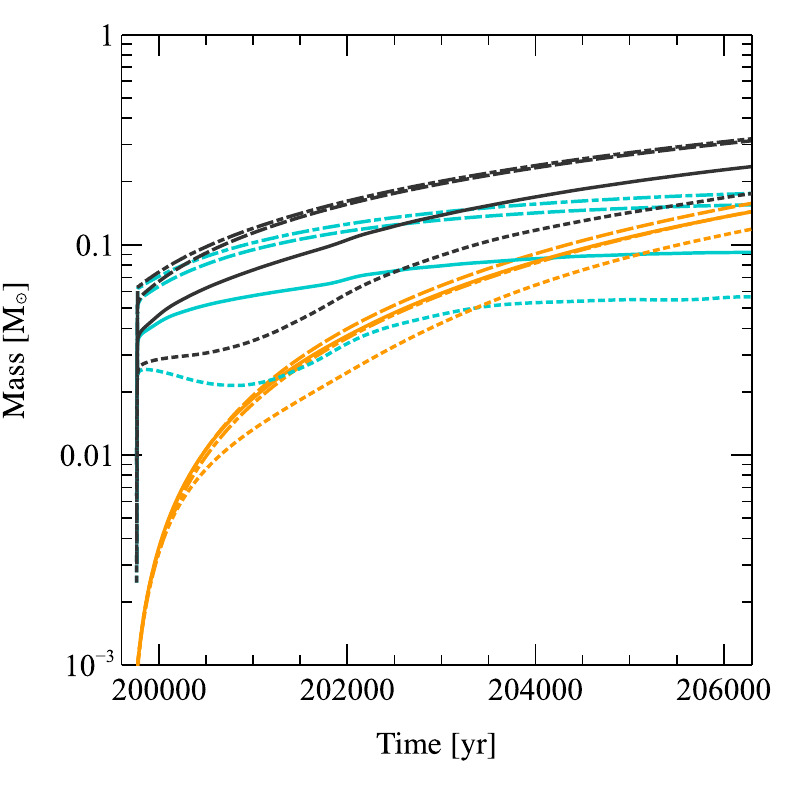}
	\includegraphics{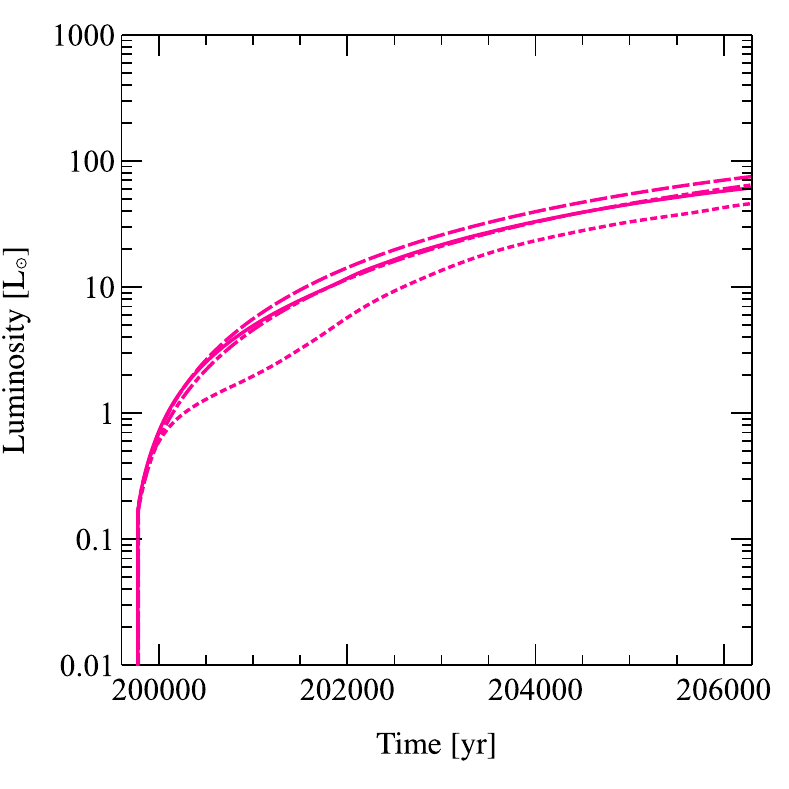}
	\caption{The evolution of the sink, protostellar, and disc masses and luminosities in the Bonnor-Ebert sphere calculations, using sink particles with luminosities set by the disc-accretion algorithm and a range of accretion radii. In the left plot, the total sink particle masses are shown in black, and the disc and protostellar masses are shown in blue and orange respectively. In the right plot, the sink luminosity is shown in magenta. Values for the 0.05-AU, 0.5-AU, 5-AU and 50-AU calculations are shown by the dotted, solid, dashed, and dot-dashed lines respectively. Larger sink particles encompass more material and have higher initial masses. The protostellar feedback heats material in the inner disc, which remains bound within larger sink particles, but expands as an outflow when using smaller sink particles, reducing their accretion rates. The lower accretion rate also reduces the protostellar luminosity.}
	\label{masslumi_fb_0001_comp}
\end{figure*}

\subsection{Disc-accretion feedback model} \label{resultsBE_disc}
Fig. \ref{denstemp_fb_0001_05AU} shows snapshots of the gas column density and temperature in a Bonnor-Ebert sphere calculation using 0.5-AU sink particles with luminosities calculated by the disc-accretion algorithm, compared to a calculation that is identical except that it uses a sink particle with no luminosity.

Similar to the calculations using constant sink luminosities, once a sink particle is inserted, the additional luminosity heats the inner region of the surrounding disc, causing it to expand. However, unlike the calculation with the strongest feedback that is described in Section \ref{resultsBE_cons}, the expansion of the disc is steady, and there is no subsequent re-collapse.

Fig. \ref{masslumi_fb_0001_05AU} shows the time evolution of the sink particle masses and luminosity in the calculation with feedback. The initial luminosity of the sink is low, beginning at $\sim 0.01 \text{L}_{\odot}$, but it reaches $\approx 2 \text{L}_{\odot}$ after 500 years. From the calculations using constant sink luminosities, it is clear that this is in the luminosity regime that does not significantly disrupt the disc structure. As such, the initial expansion of the disc following the insertion of the sink particle is small.

The luminosity steadily increases throughout the calculation, reaching a maximum value of $\sim 60 \text{L}_{\odot}$ by the end of the calculation. This is substantially higher than the $\text{10.0~L}_{\odot}$ case presented in the previous section. However, as the luminosity increases slowly from an initially low value and the sink particle mass grows with time, the rate at which the disc is heated is never high enough to cause a rapid, thermally-driven expansion. Instead, as the disc temperature continues to rise, it steadily expands but remains in hydrostatic equilibrium.

The result is a disc that is approximately twice as large and noticeably more diffuse than in the no-feedback calculation, with a typical gas density of $\sim 100~\text{g~cm}^{-2}$, compared to a typical gas surface density of $\sim 500~\text{g~cm}^{-2}$ in the no feedback calculation. The disc is also substantially hotter, with a large portion of the disc exhibiting temperatures in excess of 500 K, compared to typical temperatures of 200 K in the no-feedback calculations. 

In addition to its impact upon the surrounding disc structure, the inclusion of sink feedback also affects the evolution of the sink particle mass. The accretion rate of the sink particle with radiative feedback is lower than the sink particle with no luminosity for the first $\sim 4000$ years of the calculation. This results a sink mass which is $\approx 0.04~\text{M}_{\odot}$ lower in the feedback case than in the non-feedback case by the end of the calculations ($0.22~\text{M}_{\odot}$ versus $0.26~\text{M}_{\odot}$, respectively).

During the same period, the growth rate of the protostellar mass is marginally lower than the sink accretion rate, causing the disc mass to increase slowly. Eventually however, the disc mass reaches an approximately constant value of $0.09~\text{M}_{\odot}$, indicating that a steady-state has been reached, with the sink accretion rate equal to the protostellar accretion rate. The protostellar mass provides the dominant contribution of the total sink mass $\approx 4000~\text{yr}$ after the sink particle is inserted. When the calculation is stopped, the protostellar is $\sim 0.15~\text{M}_{\odot}$, which is 2/3 of the total sink mass of $0.22~\text{M}_{\odot}$. 

\begin{figure*}
	\includegraphics{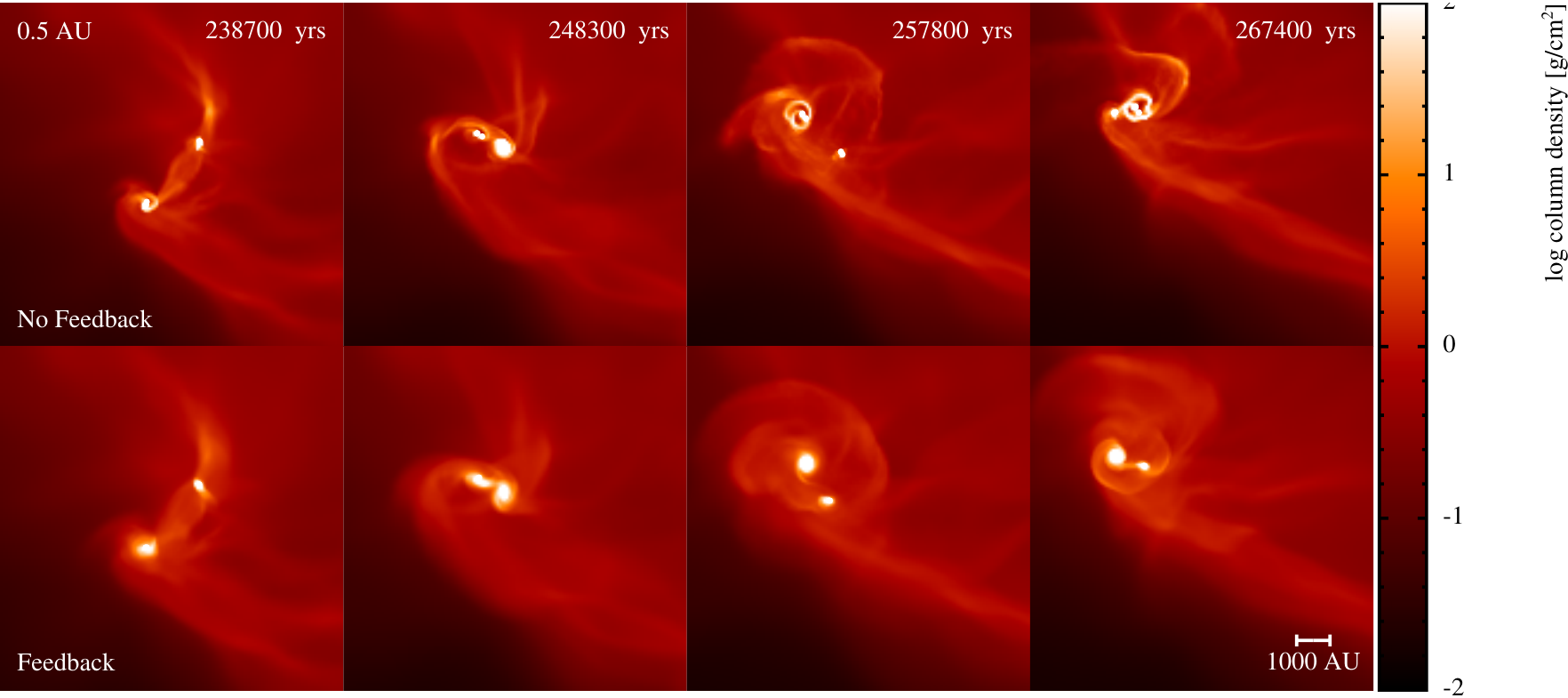}
	\includegraphics{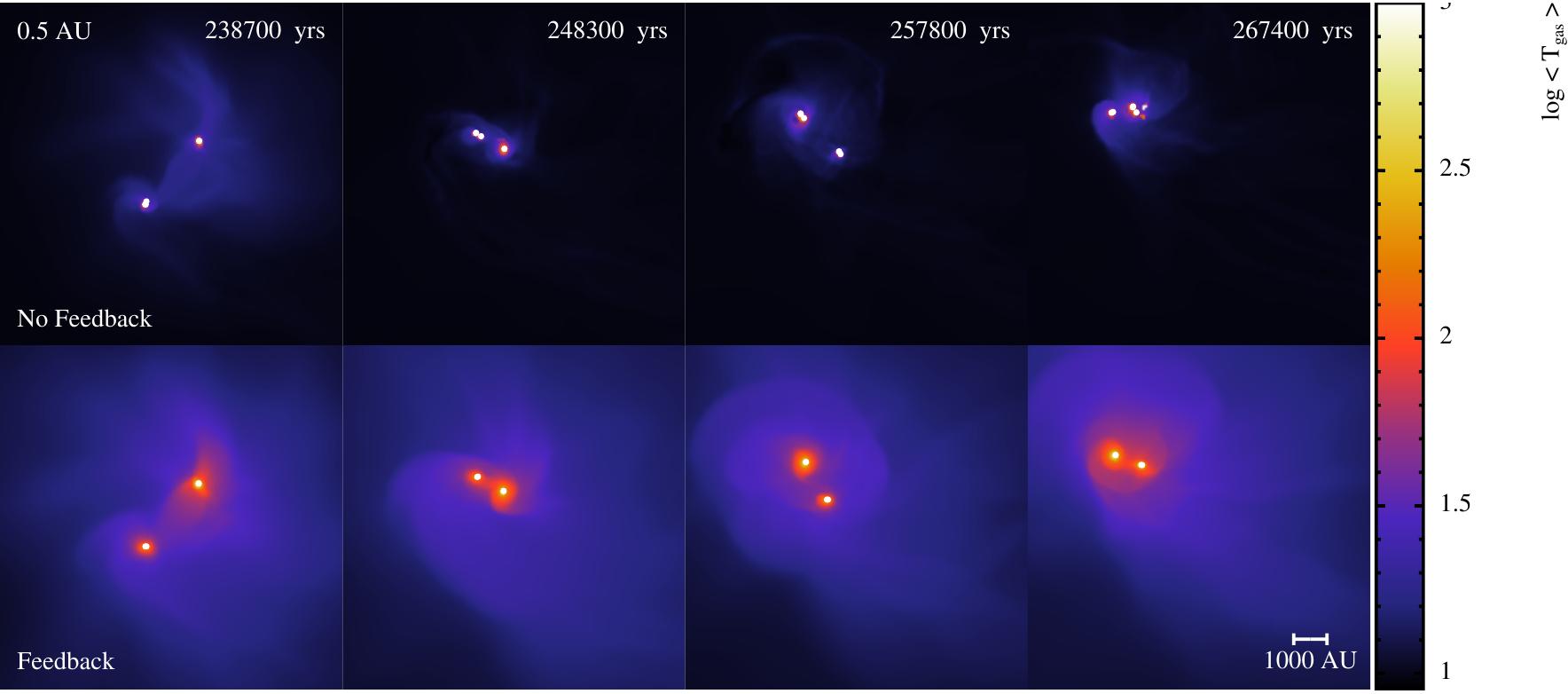}
	\caption{The time evolution of the gas column density and mass-weighted temperature in the cluster calculations using a sink particle accretion radius of 0.5 AU. The panels display the calculations at times separated by intervals of 9,549 yr, which corresponds to 1/20th of the initial cloud free-fall time. The colour scales show column densities on a logarithmic scale between $0.01~\text{g~cm}^{-2}$ and $100~\text{g~cm}^{-2}$, viewed parallel to the initial cloud's rotation axis, and temperatures on a logarithmic scale between 9 K and 1000 K. The length scale is shown in the bottom-right of each set of panels. Including radiative feedback from sinks results in higher temperatures and a more diffuse structure than in the calculation not including sink feedback, and inhibits fragmentation.}
	\label{denstemp_fb_clus_05AU}
\end{figure*}

\begin{figure*}
	\includegraphics{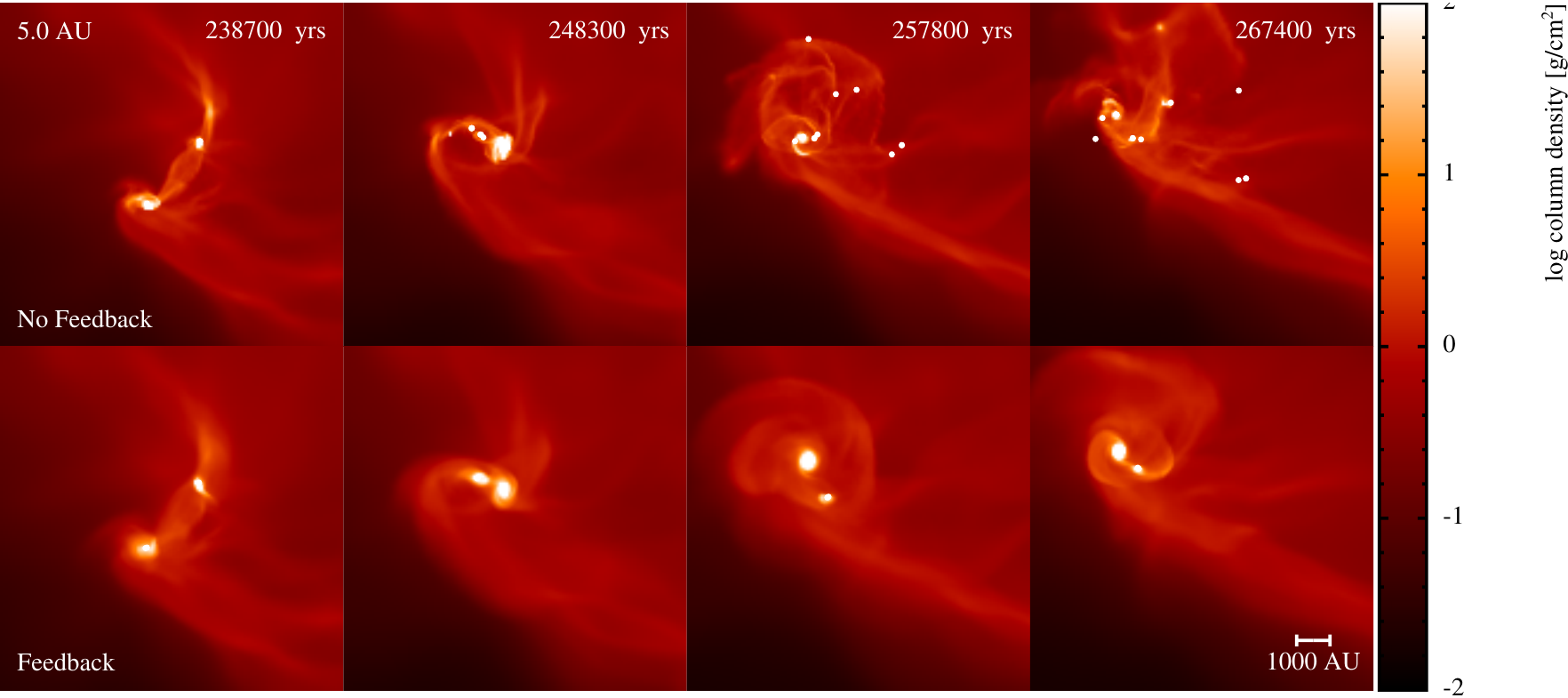}
	\includegraphics{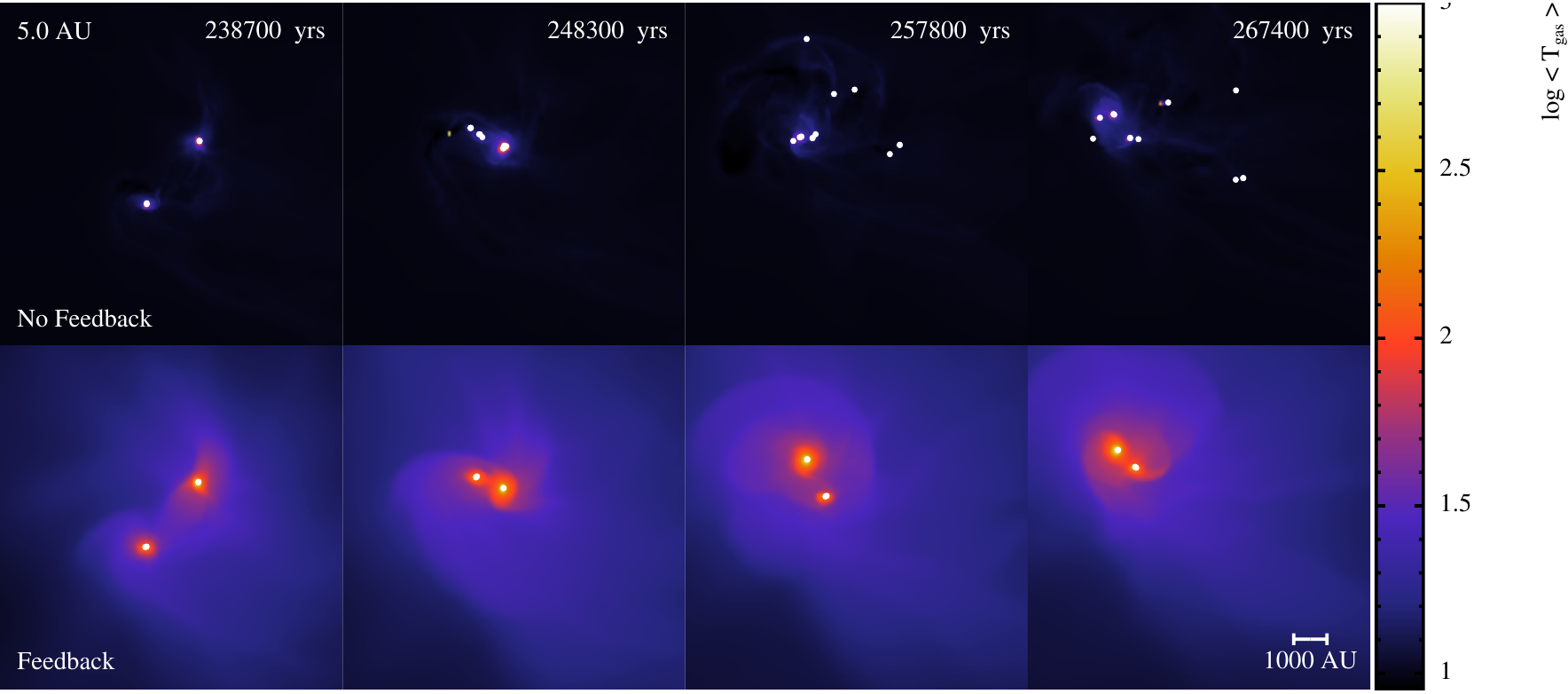}
	\caption{
	The time evolution of the gas column density and mass-weighted temperature in the cluster calculations using a sink particle accretion radius of 5 AU. The panels display the calculations at times separated by intervals of 9,549 yr, which corresponds to 1/20th of the initial cloud free-fall time. The colour scales show column densities on a logarithmic scale between $0.01~\text{g~cm}^{-2}$ and $100~\text{g~cm}^{-2}$, viewed parallel to the initial cloud's rotation axis, and temperatures on a logarithmic scale between 9 K and 1000 K. The length scale is shown in the bottom-right of each set of panels. Including radiative feedback from sinks once again results in higher temperatures and more diffuse structures than in the calculation not including sink feedback. The inhibition of fragmentation is even more evident, with the no feedback calculation producing many more objects than the calculation including feedback.}
	\label{denstemp_fb_clus_50AU}
\end{figure*}

\begin{figure}
	\includegraphics{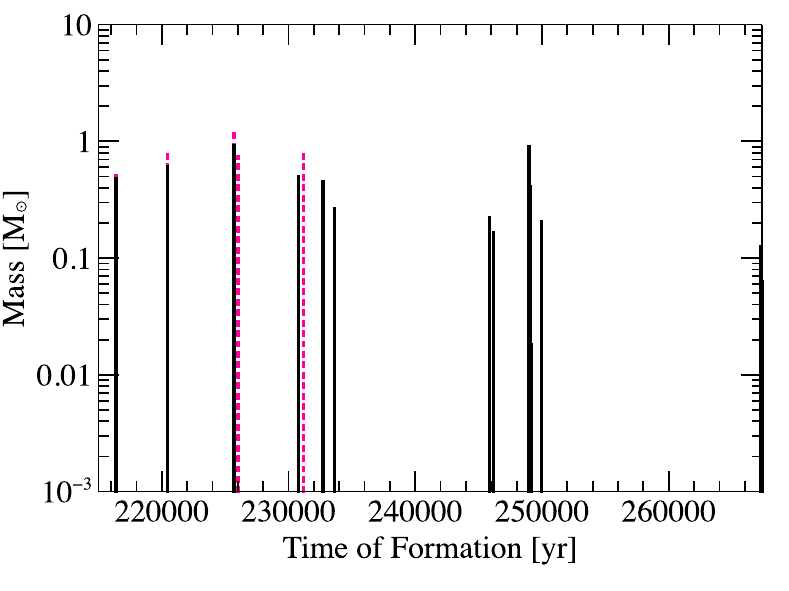}
	\includegraphics{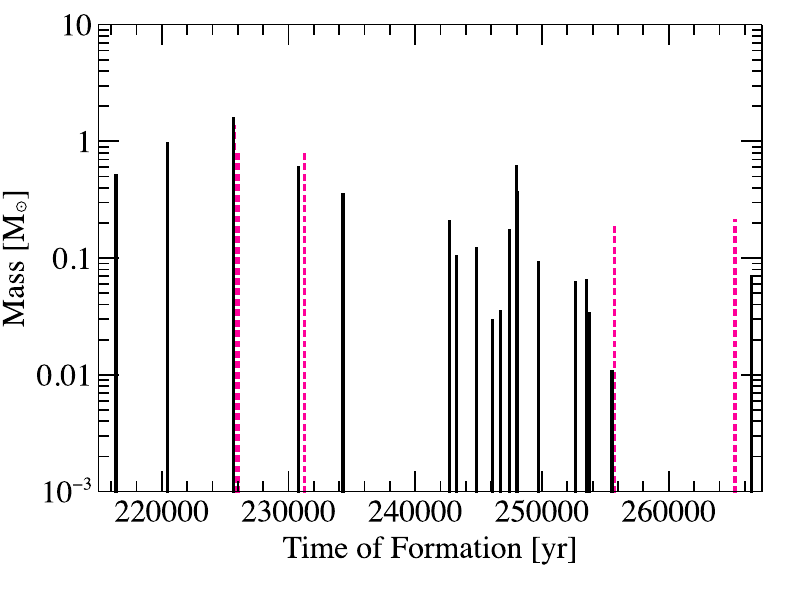}
	\caption{The mass and formation time of each star and brown dwarf in the 0.5 AU (top) and 5 AU (bottom) cluster calculations. Objects formed in the calculations including radiative feedback from sink particles are shown in magenta (dotted lines), and objects formed in the calculations without feedback from sinks are shown in black (solid lines). A similar number of objects are produced by both the feedback and no-feedback calculations early in the star-formation process. Once the feedback from the first protostars begins to heat the gas however, the gas becomes more stable against fragmentation, and fewer objects are produced than in the calculations not including feedback from sinks.}
	\label{formtime_all}
\end{figure}

\subsection{Dependence on sink particle radius} \label{resultsBE_rad}
Figure \ref{dens_fb_0001_comp} shows the gas column density in calculations using the disc-accretion algorithm to calculate the luminosities of sink particles with a range of accretion radii. The calculations using accretion radii of 0.05 AU, 0.5 AU and 5 AU produce circumstellar discs approximately 200 AU, 130 AU and 30 AU in diameter, respectively, and the 50 AU calculation does not form a resolved disc. We also see that the 0.05 AU sink particle produces a thermally-driven  outflow after being inserted (second and third panels in the top row of Fig.~\ref{dens_fb_0001_comp}).

The main differences seen in Fig.~\ref{dens_fb_0001_comp} come about from the fact that sink particles bind material inside the accretion radius once accreted and do not let it out again. As we saw in Sections \ref{resultsBE_cons} and \ref{resultsBE_disc}, the radiative feedback can cause disc expansion and thermally-driven of outflows. This can significantly alter the structure and dynamics of the surrounding disc. When the accretion radius is small, gas close to the sink particle is not bound within the sink particle. It is therefore able to expand, potentially driving outflows such as the one observed in the 0.05 AU calculation \citep[see][]{2010MNRAS.404L..79B,2011MNRAS.417.2036B}. However, when the accretion radius is large, this mass is bound within the sink particle, and no expansion is possible. 

This is illustrated by Fig. \ref{masslumi_fb_0001_comp}, which shows the evolution of the sink, disc and stellar masses, and sink luminosity in each of the calculations. There is a decrease in the rate of growth of the 0.05 AU sink for the first $\approx 1000~\text{yr}$ after its formation due to the thermally driven expansion of the inner disc. The protostellar accretion rate and luminosity are also affected. This mechanism is also present in the 0.5 AU calculation, although the effects are heavily suppressed, however it is completely absent in the 5 AU and 50 AU calculations, due to the material being bound inside the accretion radius.

Furthermore, if the sink accretion radius is large enough, it may encompass the entire first core, and much of the material that forms a disc when using smaller sink particles. This is reflected in the higher initial masses of the sink particles in Fig. \ref{masslumi_fb_0001_comp}. In this case, most of the gas surrounding the newly-formed protostar is either accreted immediately by the sink particle, or shortly after its formation, and no disc is resolved. This is particularly evident in the 50 AU calculation. 

\begin{figure*}
	\includegraphics{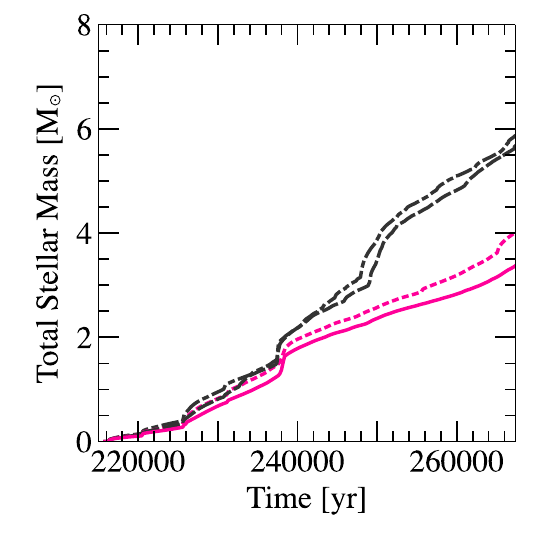}
	\includegraphics{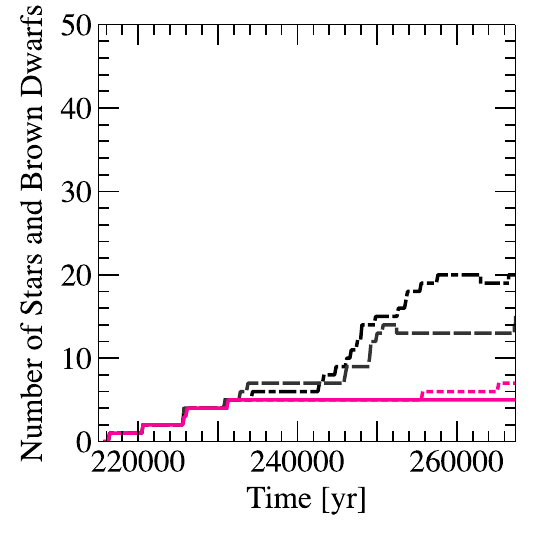}
	\includegraphics{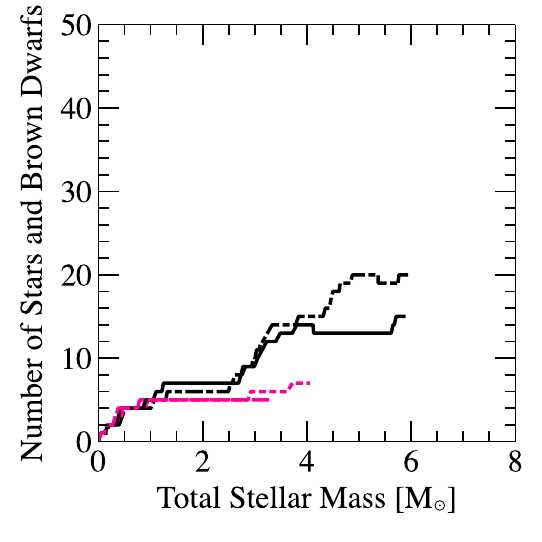}
	\caption{The evolution of the total stellar mass and total number of objects in each of the cluster calculations. The values for the calculations including sink feedback are shown in magenta (0.5 AU and 5.0 AU by solid and dotted lines, respectively), and the calculations that do not include feedback are shown in black 0.5 AU and 5.0 AU by dashed and dot-dashed, respectively). Including radiative feedback from sinks results in a lower star-formation rate, and a low number of objects produced. The total stellar mass is also distributed between fewer objects than in the calculations that do not include sink feedback.}
	\label{total_all}
\end{figure*}

\subsection{Summary}
The calculations presented here illustrate several key points about radiative feedback from sink particles, and its effects on protostellar systems. Firstly, due to the low binding energy of newly-formed sink particles and their surrounding discs, the discs are susceptible to becoming unbound and expanding due to excess thermal pressure. This primarily occurs if the luminosity of the sink particle is large at early times. We have shown that this problem may be avoided by increasing the luminosity with time from an initially small value using the disc-accretion algorithm. This gives a stable disc, but still allows the sink particle to have large luminosities a few thousand years after sink particle insertion. Finally, we have found that there is some variation in the results when using sink particles with different accretion radii. These variations mainly come about from the ability of gas at small radii to expand when using small sink particles, but that gas being trapped within the sink particle when larger accretion radii are used.  This effect is not caused by the sink feedback mechanism; it is simply a consequence of the increased resolution when using smaller sink particles. It also indicates that, regardless of what feedback prescription is used, there is always likely to be uncertainty at the level of a factor of 2--3 in the luminosity because of the uncertainty in what is happening inside the sink particle accretion radius.

\begin{figure}
	\includegraphics{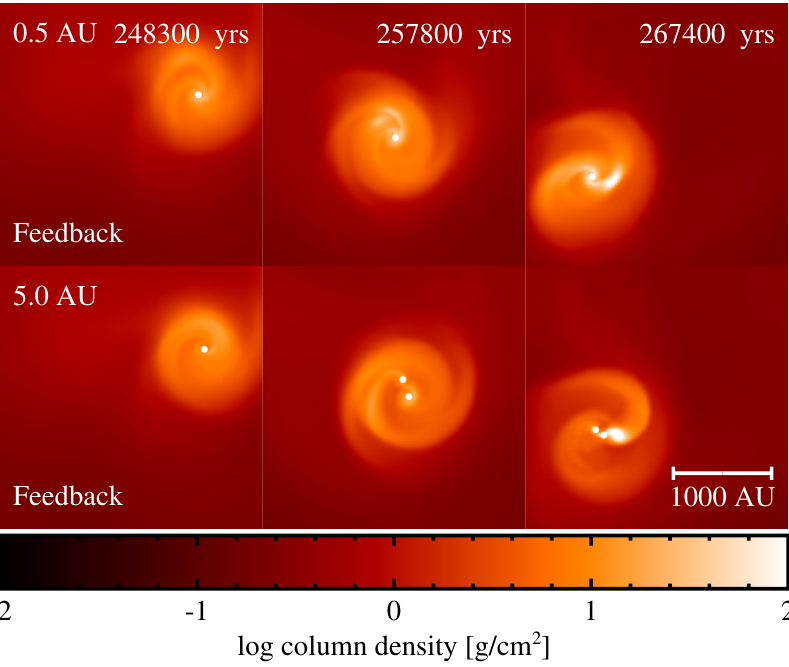}
	\includegraphics{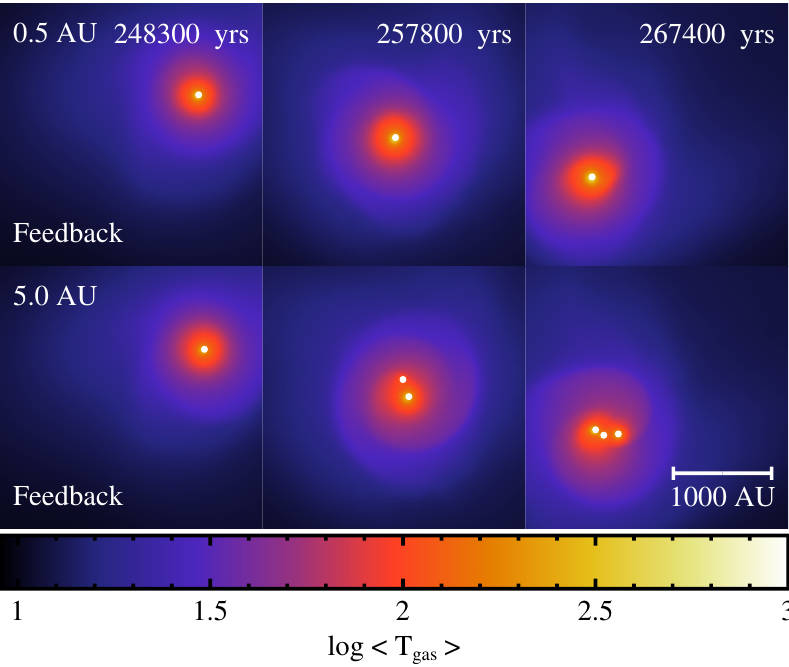}
	\caption{Fragmentation in the second star-forming core in the cluster calculations, using sink particle accretion radii of 0.5 AU and 5 AU, and including sink radiative feedback. The panels display the calculations at times separated by intervals of 9,549 yr, which corresponds to 1/20th of the initial cloud free-fall time. The colour scales show column densities on a logarithmic scale between $0.01~\text{g~cm}^{-2}$ and $100~\text{g~cm}^{-2}$, viewed parallel to the initial cloud's rotation axis, and temperatures on a logarithmic scale between 9 K and 1000 K. The length scale is shown in the bottom-right of each set of panels. In both calculations, the circumstellar discs exhibit spiral density structure. The disk fragments into three objects in the 5 AU calculation, but does not fragment in the 0.5 AU calculation. The temperature profiles are similar in both cases.}
	\label{denstemp_fb_clus_comp}
\end{figure}

\section{Small stellar clusters} \label{resultsclus}
In this section, we present results from four radiation hydrodynamical simulations that produce small clusters of stars. We compare the effects of including radiative feedback from sink particles with accretion radii of 0.5 AU and 5 AU with calculations that do not include sink feedback. Each of the calculations were run until they reached 1.40 initial cloud free-fall times. A total of 5 and 15 objects were formed by the feedback and no-feedback calculations respectively, using accretion radii of 0.5 AU, and 7 and 20 objects were formed by the feedback and no-feedback calculations respectively, using accretion radii of 5 AU. In each of the no-feedback calculations, one sink merger occurred, at times of 252,600 yr and 263,200 yr in the 0.5 AU and 5 AU calculations, respectively. 

Section \ref{resultsclus_evo} describes the time evolution of the clouds, including the formation of objects and their impact on the surrounding cloud structure. In Section \ref{resultsclus_IMF}, we examine the distribution of masses in each cluster, and compare the results from the calculations including sink feedback to those ignoring feedback in order to probe its effects on the statistical properties of clusters.

\subsection{The evolution of the clouds} \label{resultsclus_evo}
Figs.~\ref{denstemp_fb_clus_05AU} and \ref{denstemp_fb_clus_50AU} show the evolution of the gas density and temperature in the main star-forming core during each of the cluster calculations.  A few protostars also form in other regions of the cloud, but we only show a single dense core here for clarity. In each of the clusters, the initial `turbulent' velocity field results in a filamentary structure. Where flows of gas collide, the turbulent energy is dissipated by shocks forming filaments.  The interactions of these filaments produce dense cores that can collapse to produce stars, often at the intersections of filaments. Star formation occurs primarily in these cores, which heat the surrounding region as gas falls into the gravitational potential and is accreted by newly formed protostars. 

Including radiative feedback from sink particles results in significantly different evolution of the clusters. Once the first protostars form, the feedback generated by accretion raises the temperature of the surrounding gas. This results in a hotter, more diffuse density structure that is less prone to fragmentation. Contrastingly, in the calculations that do not include sink feedback, the gas is much cooler and therefore more unstable to collapse. This leads to the fragmentation of several massive discs that form around the first protostars.

Fig. \ref{formtime_all}, which shows the formation time and final mass of each star and brown dwarf in the cluster calculations, reflects this. The formation times of the first four objects are almost identical in both the feedback and no-feedback calculations. However, as the calculations progress, those that do not include feedback from sink particles begin to rapidly form more objects due to fragmentation of the gas. This is particularly true for the 5 AU no-feedback calculation, which produces many more objects than the 0.5 AU no-feedback calculation due to the lack of heating from material being resolved at small radii. However, in the calculations including sink feedback, the accretion luminosity dominates, and is calculated regardless of the accretion radius. This results in highly similar temperature distributions, and therefore similar stellar populations.

Fig. \ref{total_all} shows the evolution of the total stellar mass (that is, the combined mass of the sink particles) and total number of objects in each of the calculations. The difference in the rate of fragmentation between the calculations is even more obvious here. The total mass and number of objects are similar across all of the calculations for the first $\sim 20,000$ years. However, in the calculations including sink feedback, once the luminosity from the first protostars begins to heat the gas, the rate of star-formation becomes heavily suppressed, reducing the number of objects formed by 1/2 to 2/3, depending the accretion radii by the end of the calculations. The total stellar mass is also reduced by $\approx 40$\% by the end of the calculations when including feedback, due to the additional support against collapse provided by the extra thermal pressure generated from the extra luminosity, which decreases the average protostellar accretion rate. This is similar to the findings of \citet{2012MNRAS.419.3115B} when comparing calculations using a barotropic equation of state with calculations using radiative transfer but without luminous sink particles.  Note that there is very little dependence of the total stellar mass on whether 0.5 or 5-AU sink particles are used; what little difference there is can be attributed to the fact that the latter includes mass lying between 0.5 and 5 AU of a sink particle, while with the smaller accretion radii this mass is still part of the hydrodynamical simulation and so is not counted.

Plotting the number of objects against the total stellar mass illustrates how the distribution of mass changes throughout the duration of the calculations. Whilst in the calculations that do not include sink feedback the number of objects increases steadily with the total stellar mass, in the feedback calculations, after an initial period of star formation, the number of objects remains almost constant as the stellar mass increases. This reflects the lack of fragmentation in the calculations including sink feedback. The result is that the first protostars to form are able to accrete from the available gas reservoir with little competition from other nearby objects. As such, they account for a larger proportion of the total stellar mass. They are also able to attain higher masses than their counterparts in the calculations that do not include sink feedback, despite the lower total stellar mass in the feedback calculations.

Importantly, when feedback from the sink particles is included, there is very little dependence of the results on whether 0.5 or 5-AU sink particles are used.  In Fig.~\ref{total_all}, the evolution of the total stellar mass is very similar (and the small difference can be attributed to the difference in the way mass between 0.5 and 5~AU is treated). The numbers of protostars formed and the times at which they form are also similar. The two additional objects formed towards the end of the 5 AU calculation are due to the fragmentation of a massive disc in the second star-forming core, shown in Fig. \ref{denstemp_fb_clus_comp}. The second panel shows that the density structure of the discs in the 0.5 AU and 5 AU calculations are highly similar, however the 5 AU disc fragments, whereas the 0.5 AU disc does not. Given the similarity between the two systems, we conclude that although the feedback algorithm does remove most of the dependence of the calculations on the value of the sink particle accretion radius, small differences may still affect disc fragmentation. This lack of dependence on sink particle radius is exactly what we would hope for from a `sub-grid' model.  By contrast, when only including radiative feedback from resolved gas, although reducing the sink particle accretion radius does result in less fragmentation, the number of protostars formed in the calculation using 0.5 AU is still a long way from converging to the result obtained with sink particle feedback (twice as many protostars are formed).  

Fig. \ref{proto_all} shows the evolution of the sink masses and luminosities in each of the calculations including feedback. The masses of the sink particles increase rapidly during the period shortly after their formation, as they accrete the remains of the first hydrostatic cores. Following this period of rapid accretion, the masses continue to increase at a lower rate that is relatively stable. There is a spike at approximately 238,000 years in the mass accretion rates of two sink particles. Comparing with Fig. \ref{denstemp_fb_clus_05AU} and Fig. \ref{denstemp_fb_clus_50AU}, we see that this is due to a dynamical interaction between two protostellar discs that removes angular momentum from the disc material, increasing the accretion rates. 

Whilst the masses increase at a reasonably steady rate for the majority of the calculations, the luminosities of the sinks are more variable. After the initial phase of rapid accretion, there is a general trend towards higher luminosities, reflecting the increasing mass of the protostars. However, within this trend there is significant variation. The spike in the accretion rate of two sink particles due to the dynamic interaction of their discs is noticeable, leading the higher mass particle to temporarily reach $500~\text{L}_{\odot}$. At the end of the calculations, the luminosities of the first five protostars to form exhibit a spread of approximately one order of magnitude, compared to the masses, which only vary by a factor of $\sim 2$.  These variations are primarily caused by changes in the protostellar accretion rates. This is clear from the third panel of the figure, which shows that the luminosities of several sink particles decrease despite their masses increasing. Changes in the accretion rates are largely due to dynamics and filamentary structures in the cluster gas, which are produced by the turbulent velocity field and/or dynamical encounters.

\subsection{The protostellar mass function} \label{resultsclus_IMF}
\begin{figure}
	\includegraphics{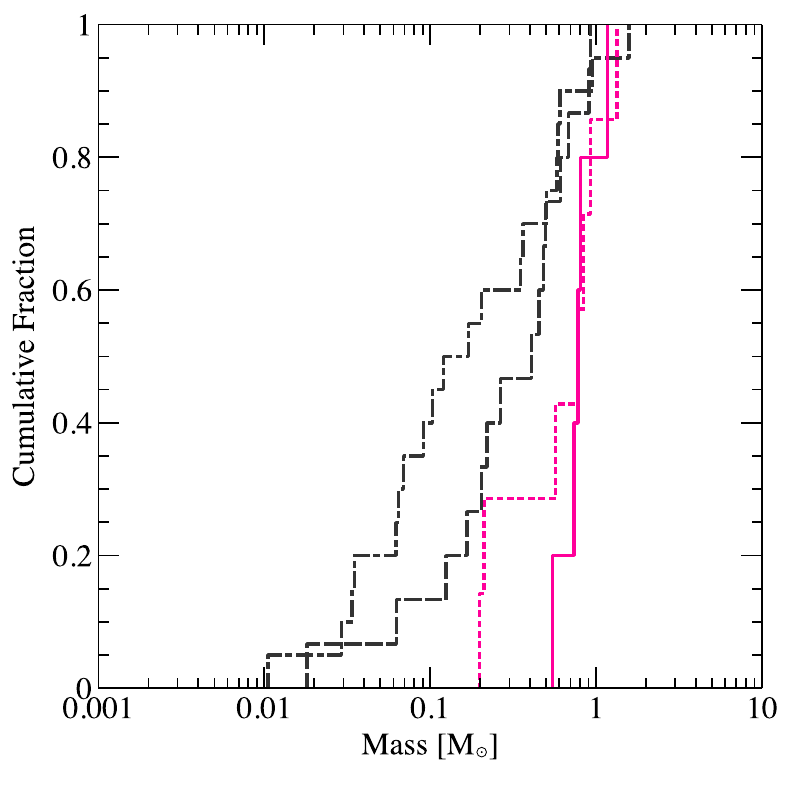}
	\caption{The cumulative distribution of stellar and brown dwarf masses at the end of each of the calculations. The values for the calculations including sink feedback are shown in magenta (0.5 AU and 5.0 AU by solid and dotted lines, respectively), and the calculations that do not include feedback are shown in black 0.5 AU and 5.0 AU by dashed and dot-dashed lines, respectively). Including radiative feedback from sink particles increases the median mass of the clusters. It also decreases the difference between the distributions in the calculations using accretion radii of 0.5 AU and 5 AU.}
	\label{IMF_cum_all}
\end{figure}

Fig. \ref{IMF_cum_all} shows the cumulative protostellar mass function at the end of each of the calculations ($t=1.40~t_{\rm ff}$). We do not plot the differential mass functions, as too few objects were produced. 

It is clear that including feedback from sink particles has a significant impact on the distribution of masses formed in the groups. The additional thermal pressure provided by the sink particle luminosities inhibits fragmentation, leading to a lower number of low-mass objects. With fewer objects to compete with, protostars that form are able to accrete a larger fraction of the available gas reservoir, resulting in a higher final mass. This increases the median mass of the cluster, from $0.41~\text{M}_{\odot}$ to $0.77~\text{M}_{\odot}$ in the 0.5 AU calculations, and from $0.12~\text{M}_{\odot}$ to $0.77~\text{M}_{\odot}$ in the 5 AU calculations. 

The median masses and the overall mass distributions of the calculations using sink particle feedback but different accretion radii are very similar.  This reflects the ability of the disc-accretion feedback method to account well for the luminosity lost when using different sink particle accretion radii, as also found in the previous section.  

By contrast, in the no feedback cases, the median stellar masses for the calculations without sink particle feedback differ by more than a factor of three when the accretion radius is changed from 0.5 to 5 AU.  Using an accretion radius of 5 AU results in a significant reduction in the temperature of the surrounding gas, compared to the 0.5 AU calculation, as shown in Fig. \ref{denstemp_fb_clus_05AU} and Fig. \ref{denstemp_fb_clus_50AU}. This makes the gas much more unstable to fragmentation, increasing the number of low-mass objects in the cluster relative to the 0.5 AU calculation.

\section{Discussion} \label{disc}
The results of our calculations show that including radiative feedback from sink particles in simulations of star-formation significantly affects both the dynamics of the cluster gas and the distribution of protostellar masses produced. In this Section, we consider our results in the context of previous simulations of star formation, examining the implications for the protostellar mass function and luminosity evolution.  We also compare our results to observations of protostellar heating within molecular clouds.

\subsection{Protostellar mass function} \label{disc_IMF}
Including feedback from sink particles has a significant impact the distribution of protostellar masses in star-forming cores. The reduced level of fragmentation increases the median stellar mass, as fewer objects means less competition for accretion, enabling the protostars that do form to accrete more of the reservoir and attain higher masses.

Our results follow similar patterns to those found by \citet{2009MNRAS.392.1363B,2012MNRAS.419.3115B} when comparing calculations including radiative transfer with those using a barotropic equation of state. The initial phase of star-formation in each dense core does not depend greatly on whether radiative transfer or a barotropic equation of state is used, or on whether or not sink particle feedback is included. However, once the first few protostars have formed, the additional heating begins to inhibit fragmentation and slow the rate of star formation. In addition to one large calculation, \citet{2012MNRAS.419.3115B} performed small radiation hydrodynamical star formation calculations very similar to those we present here, using 5, 0.5, and 0.05 AU non-luminous sink particles that produced 15, 11, and 8 protostars, respectively.  He also performed a barotropic version of the calculation that produced 48 protostars.  Based on these results, he argued that using radiative transfer with 0.5-AU sink particles feedback was sufficient to capture most of the effects of radiative feedback. However, while this may be true, our results with radiative feedback from sink particles show that the additional luminosity can produce roughly a factor of two reduction in the number of protostars and a corresponding factor of two increase in the median stellar mass. Calculations performed by \citet{2014MNRAS.439.3039L} including radiative feedback from sink particles also observed a similar reduction in the number of objects formed. These results suggest that the additional accretion luminosity produced inside sink particles does need to be accounted for in calculations.

We may also consider the typical size of regions that collapse to form protostars. The mass of these regions should be approximately the same as the typical protostellar mass, which can be identified with the median mass of the cluster. In an isothermal system, this is described by the typical Jeans length which only depends on the density and initial temperature. However, the luminosity from an embedded protostar will heat the surrounding gas, increasing the size of the region required to overcome the thermal pressure of the gas. \citet{2009MNRAS.392.1363B} described this as the `effective Jeans length', and identified the corresponding `effective Jeans mass'. 

By considering simple analytical arguments, \citet{2009MNRAS.392.1363B} predicted that the median mass should scale as $M_c \propto \rho^{-1/5} L^{3/10}$ where $\rho$ is the density of the gas, and $L$ is the protostellar luminosity. As the accretion luminosity scales as $L \propto 1/R_*$, our no-feedback calculations using sink particles with accretion radii of 0.5 AU and 5 AU under estimate the luminosity by factors of 50 and 500 respectively for a typical protostellar radius of $2~{\rm R}_{\odot}$ \citep{1969MNRAS.145..271L}. Thus, the analytical model of \citet{2009MNRAS.392.1363B} predicts that this should result in a shift in the median mass by factors of 3.2 and 6.5 respectively.  Given the small numbers of protostars formed, this is in good agreement with our results (using 5 AU sink particles, the median mass increases from 0.12 to 0.77~M$_\odot$, a factor of 6.4, and using 0.5 AU sink particles, the median mass increased from 0.41 to 0.77~M$_\odot$, a factor of 1.9).

We also find that the mass distributions produced by the calculations including sink particle feedback are almost independent of whether accretion radii of 0.5 AU or 5 AU are used. This agrees with the assertion of \citet{2009ApJ...703..131O} that the protostellar luminosity becomes the dominant source of heating once the first protostars have formed, as it implies that the primary source of heating is included, regardless of the sink particle accretion radius.

However, the median masses of the clusters in the feedback calculations are higher than typically observed in local star-forming regions \citep{2010ARA&amp;A..48..339B} and in the canonical Galactic stellar initial mass function \citep{2005ASSL..327...41C}. This was also true of calculations performed by \citet{2009MNRAS.392.1363B}, which used similar initial conditions to those we use in this paper, and those of \citep{2009ApJ...703..131O}. The star formation in our calculations and in these earlier calculations occurred in small protostellar groups. \citet{2018MNRAS.478.2650J} found that low-density molecular clouds form stars in well separated groups, similar to the groups that form in our calculations and the calculations of \citet{2009MNRAS.392.1363B} and \citet{2009ApJ...703..131O}. These low-density clusters were also found to exhibit higher median stellar masses than those obtained from higher density star-forming regions. \citet{2004A&amp;A...419..543G} suggested that this may be caused by ejections of low-mass objects from the groups, which decrease the number of accreting objects whilst depleting the mass of the group by a negligible amount. In fact, our calculations suggest that it is in fact due to a lack of dynamical interactions, due to the small numbers of stars within each group. With fewer dynamical interactions and ejections, most objects are able to accrete to higher masses without having their growth terminated prematurely, raising the median mass.  The implication is that to obtain the typical Galactic stellar mass function with a lower median stellar mass, stars must form in groups that are large enough and dynamically interactive enough to fully populate the low-mass end of the mass function.

It is also possible that the feedback from first protostars in the group may inhibit fragmentation long enough to allow them to accrete a substantial amount of mass before any new protostars form. Once new objects do form, the larger masses of the existing protostars allows them to competitively accrete within the small group, attaining  a higher mass than would otherwise be expected. Simulations of star formation in massive protostellar cores performed by \citet{2007ApJ...656..959K} displayed similar behaviour, with the feedback from the first protostar to form inhibiting fragmentation in the group, allowing it to accrete the majority of the available mass.

A possibility to lower the median stellar mass is that disc accretion is episodic rather than continuous. \citet{2014MNRAS.439.3039L} considered the effects of no feedback, constant feedback and episodic feedback for protostars formed in an ensemble of star-forming cores. As in our calculations, they found that continuous feedback heavily suppressed fragmentation, causing low-mass objects to be underproduced, resulting in a higher median mass. However, when including episodic feedback from protostars, this issue was resolved. In the episodic case, they varied protostellar luminosities from $\sim 0.1~{\rm L}_{\odot}$ during quiescent periods to $\sim 100~{\rm L}_{\odot}$ during outbursts. Due to the short timescales over which outbursts occurred, the distributions of protostellar masses in these calculations were highly similar to the calculations that did not include protostellar feedback. \citet{2015MNRAS.447.1550L} further showed that calculations using an episodic model of accretion were able to produce small stellar clusters whose multiple system properties were in good agreement with observations. The main \textbf{caveat} with this solution however, is that the characteristics of episodic accretion are poorly understood.

\begin{figure}
	\includegraphics{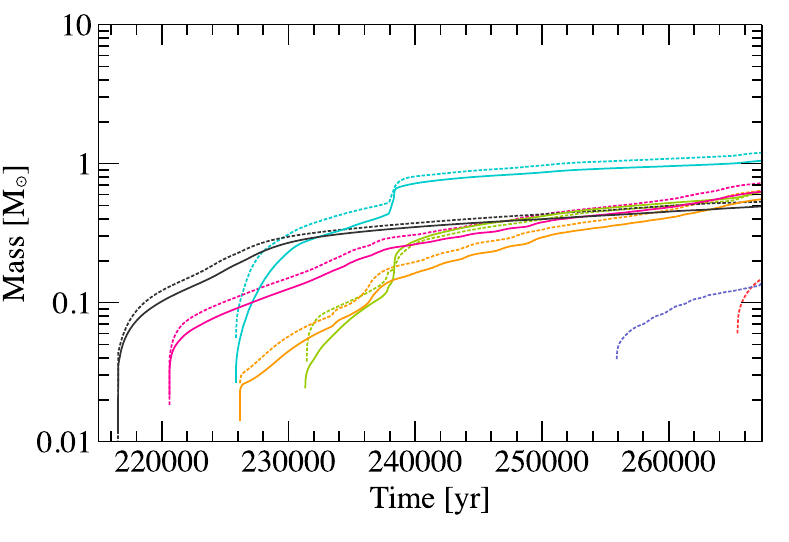}
	\includegraphics{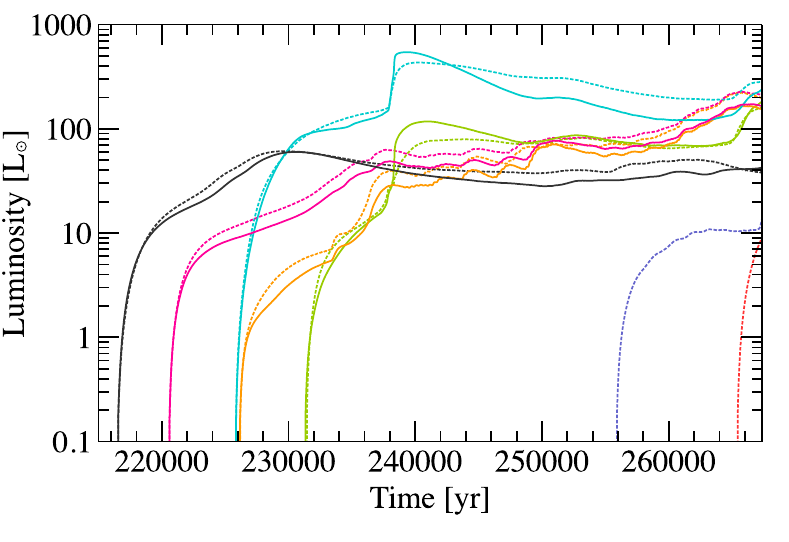}
	\includegraphics{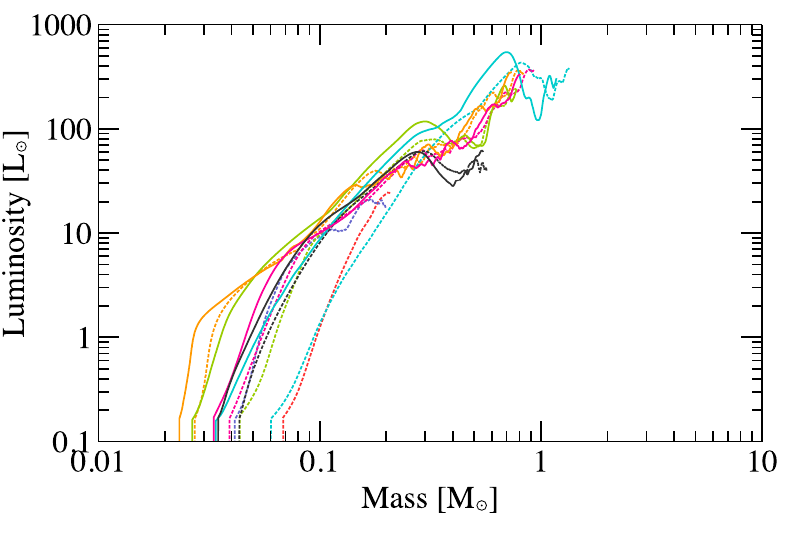}
	\caption{The evolution of the masses and luminosities and sink particles in the feedback calculations. Each colour denotes a different sink particle. The results of the 0.5 AU calculations are shown by the solid lines, and the 5 AU calculations are shown by the dotted lines. Each colour denotes a different sink particle. Once formed, the masses of the protostars increase steadily as the calculations progress. The protostellar luminosities increase as the masses increase. However, there is also considerable spread in luminosities of sink particles with similar masses, due to variations in the accretion rates.}
	\label{proto_all}
\end{figure}

\subsection{Protostellar evolution} \label{disc_proto}
Fig. \ref{formtime_all} indicates that there is little correlation between the final masses and the formation time of protostars, unlike  simple predictions \citep[e.g.][]{1977ApJ...214..488S}. Instead the final protostellar masses are determined by their accretion histories, which are largely governed through competitive accretion, in agreement with the results of \citet{1997MNRAS.285..201B,2001MNRAS.323..785B}, \citet*{1998ApJ...501L.205K}, \citet{2001ApJ...556..837K}, and \citep{2005MNRAS.356.1201B}.

The average luminosities of the protostars increase with final mass, in agreement with the findings of \citet{2010ApJ...710.1343U}. However, there is significant spread among the luminosities of protostars with similar masses, as seen in Fig. \ref{proto_all}. This is because the luminosities of protostars depend not only on their mass, but also on variations in their accretion rates due to the dynamics of their environments. During the formation of the first protostars, the accretion rates and luminosities increase smoothly. However, once more objects form, the increased frequency of dynamic interactions causes variations in the accretion rates, which give rise to variations in the accretion luminosities. 

The luminosities typically differ by less than an order of magnitude after the initial growth phase ($M_* \gtrsim 0.1~{\rm M}_\odot$).  Prior to this the luminosities may increase by up to two orders of magnitude as the surrounding remnants of their preceding first hydrostatic cores are accreted. This is in broad agreement with the results of \citet{2009ApJ...703..131O}, who found that accretion rates were smoother once feedback was introduced. However, the sink particle accretion radii used in those calculations were $30-250$ times larger than the smallest radius used here. This suggests that the dynamics driving these luminosity variations on 100-1000~yr timescales have similar effects at $\approx 1~\text{AU}$ and $\approx 100~\text{AU}$ scales. Contrastingly, \citet{2010ApJ...710.1343U} found that their protostellar accretion rates were highly variable. The reason for this is unclear.  It is possible that their method for calculating protostellar luminosities, via interpolation over the protostellar mass and accretion rate, may not fully resolve the self-regulating effect of the luminosity on the accretion flow, and may therefore over-estimate its variability. 

The typical size of these luminosity variations is much smaller than those employed in episodic models of accretion feedback, such as those of \citet{2015MNRAS.447.1550L} which vary by three orders of magnitude. If such episodic accretion does occur, it must come from some type of disc instability that is not captured by the physical processes modelled in our simulations.

\subsection{Comparison with observations}
Observational studies of young stellar objects (YSOs) have found substantial evidence of protostellar heating in low-mass star-forming regions. \citet{2009A&amp;A...501..633V,2009A&amp;A...507.1425V} conducted observations using APEX-CHAMP to map high-$J$ CO lines around $\sim 30$ nearby sources in order to trace the distribution of warm gas. \citet{2010A&amp;A...518L.121V} subsequently performed spectroscopy using Herschel/PACS to study the spectra of these sources. Combining these data sets, both \citet{2012A&amp;A...537A..55V} and \citet{2012A&amp;A...542A..86Y} confirm that these studies indicate the presence of heated regions of $\approx 100$~K extending out to $\sim 1000$~AU in the vicinity of embedded low-mass protostars. 

\citet{2013A&amp;A...551A..34S} found similar results, using multi-wavelength Herschel observations to create temperature maps of the Corona Australis star-forming region. The temperature distributions in our calculations including protostellar feedback (see Figs. \ref{denstemp_fb_clus_05AU} and \ref{denstemp_fb_clus_50AU}) are remarkably similar to those of Sicilia-Aguilar et al. (2013), with heated regions of $\sim 30-100~\text{K}$ extending for several thousand AU around the newly formed protostars. 

\citet{2013MNRAS.429L..10H} used James Clerk Maxwell Telescope (JCMT) SCUBA-2 observations to map the heating from young stars and protostars in the NGC 1333 star-forming region. They also detected elevated temperatures upwards of $20~\text{K}$ in the vicinity of embedded protostars, and argued that this heating effect would lead to an increase in the typical mass of the next generation of protostars as we find in our simulations. Similar temperature profiles were observed by \citet{2016ApJ...826...95C} in the vicinity of young stellar objects, using a combination of JCMT and Herschel observations to map dust temperatures in the Perseus molecular cloud.

The protostars in both our calculations and in the observations of \citet{2013A&amp;A...551A..34S}, \citet{2013MNRAS.429L..10H}, and \citet{2016ApJ...826...95C} form in small groups approximately 0.1 pc across. Similar structures have been observed in the nearby Taurus-Auriga region \citep{2003ApJ...590..348L}, which also exhibit a median stellar mass that is higher than expected. \citet{2004A&amp;A...419..543G} have suggested that the formation of stars in groups such as this may lead to increased masses, as low-mass objects are often dynamically ejected, leaving fewer objects to compete for accretion of the gas reservoir. However, the lack of brown dwarfs in our calculations suggests that this is not the case, as ejections would increase the number of objects that have their accretion terminated before reaching higher masses. Instead, it seems more probable that inhibition of fragmentation via protostellar heating, combined with a lack of dynamical ejections, is responsible for the increased median mass of stars in these regions.

\section{Conclusions} \label{conc}
In this paper, we have presented a new method for including radiative feedback from sink particles in smoothed particle hydrodynamics simulations of star formation.  We have also presented results from calculations of star formation in small clusters to investigate the effects of including sink feedback on the formation of stellar clusters and their properties.

We find that including sink feedback has a significant effect on the star formation process and the evolution of star-forming clusters. Our conclusions are as follows:
\begin{enumerate}
	\item The additional heating from newly formed protostars in calculations including radiative feedback from sink particles suppresses fragmentation even further than in calculations that include radiative transfer and use small sink particles that do not emit radiation. This reduces the star formation rate after the formation of the first protostars, resulting in a lower total number of objects formed and total stellar mass at the end of the calculations. 

	\item The reduced number of objects allows the protostars to attain higher final masses than in calculations with no sink particle feedback. The final protostellar masses are a function of their accretion histories, rather than simply depending on their formation times. 

	\item The average protostellar luminosities are found to not only to depend on the protostellar mass, but also to depend strongly on the mass accretion rate, causing an order of magnitude spread in the luminosities of protostars with similar masses. The luminosity variations of protostars are largely due to changes in the accretion rate driven by the dynamics of the cluster gas.

	\item Including feedback from sink particles raises the median mass of protostars in the calculations. The magnitude of the effect is in agreement with the scaling predicted by the simple analytical model proposed by \citet{2009MNRAS.392.1363B}. We also find that the stellar mass distributions obtained from calculations that include radiative feedback from sink particles are insensitive to the choice of sink particle accretion radii (varied between 0.5 and 5 AU). This implies that protostellar heating from accretion is the dominant form of heating in star-forming clusters, as asserted by \citet{2009ApJ...703..131O}.

	\item The median masses of the clusters in feedback calculations are higher than typically observed, due to a lack of brown dwarfs. This provides an alternative explanation for the high median mass of similar groups observed in the Taurus-Auriga region from that given by \citet{2004A&amp;A...419..543G}. They proposed that low-mass objects were preferentially ejected, leaving behind higher-mass stars.  Instead, the higher median mass may be due to a {\it lack} of dynamical interactions and ejections in small groups of protostars so that brown dwarfs are underproduced.

	\item The temperature distributions in the calculations including radiative feedback from sink particles are similar to those observed in Galactic star-forming regions, such as NGC 1333 \citep{2013MNRAS.429L..10H}, and the Corona Australis \citep{2013A&amp;A...551A..34S} and Perseus regions \citep{2016ApJ...826...95C}. 
\end{enumerate}

\section*{Acknowledgements}
This work was supported by the European Research Council under the European Commission's Seventh Framework Programme (FP7/2007-2013 Grant Agreement No. 339248). The calculations discussed in this paper were performed on the University of Exeter Supercomputer, Isca. The rendered plots shown were produced using SPLASH \citep{2007PASA...24..159P}.




\bibliographystyle{mnras}
\bibliography{feedback}


\bsp	
\label{lastpage}
\end{document}